\documentclass{article}

\usepackage[preprint]{log_2026}
\usepackage{natbib}
\usepackage[T1]{fontenc}
\usepackage{microtype}
\usepackage{graphicx}
\usepackage{booktabs}
\usepackage{array}
\usepackage{capt-of}
\usepackage{amsmath,amssymb,amsthm}
\usepackage{mathtools}
\usepackage{xcolor}
\usepackage{tikz}
\usepackage{eso-pic}
\usetikzlibrary{arrows.meta,positioning,calc}
\usepackage{algorithm}
\usepackage{algorithmic}
\usepackage[backref=page]{hyperref}
\usepackage[capitalize,noabbrev]{cleveref}

\theoremstyle{plain}
\newtheorem{theorem}{Theorem}
\newtheorem{proposition}[theorem]{Proposition}
\newtheorem{lemma}[theorem]{Lemma}
\newtheorem{corollary}[theorem]{Corollary}

\AddToHook{env/proposition/begin}{\crefalias{theorem}{proposition}}
\AddToHook{env/lemma/begin}{\crefalias{theorem}{lemma}}
\AddToHook{env/corollary/begin}{\crefalias{theorem}{corollary}}

\theoremstyle{definition}
\newtheorem{definition}{Definition}

\crefname{lemma}{Lemma}{Lemmas}
\crefname{proposition}{Proposition}{Propositions}

\newcommand{\pending}[1]{{\color{red}\ttfamily\bfseries[\detokenize{#1}]}}
\makeatletter
\DeclareRobustCommand{\res}[1]{\@ifundefined{res@#1}{\pending{#1}}{\@nameuse{res@#1}}}
\newcommand{\resrows}[1]{\@ifundefined{resrows@#1}{\multicolumn{1}{l}{\pending{rows #1}}\\}{\@nameuse{resrows@#1}}}
\let\orcut@includegraphics\includegraphics
\renewcommand{\includegraphics}[2][]{\IfFileExists{#2}{\orcut@includegraphics[#1]{#2}}{%
  \fbox{\parbox[c][2.6cm][c]{0.9\linewidth}{\centering\color{red}\ttfamily\footnotesize[missing figure: \detokenize{#2}]}}}}
\makeatother
\IfFileExists{generated/solver_numbers.tex}{\input{generated/solver_numbers.tex}}{}
\makeatletter
\def\orcut@expect#1#2{\@ifundefined{res@#1}{\global\let\orcut@stale\relax}{%
  \protected@edef\orcut@a{\@nameuse{res@#1}}\def\orcut@b{#2}%
  \ifx\orcut@a\orcut@b\else\global\let\orcut@stale\relax\fi}}
\orcut@expect{ArtifactsLayout}{onecolumn}\orcut@expect{ArtifactsMethod}{orcut}
\AtBeginDocument{\ifdefined\orcut@stale\AddToShipoutPictureFG*{\put(\LenToUnit{0.6in},\LenToUnit{0.45in}){%
  \color{red}\bfseries\footnotesize generated/ was not written for this version: run experiments/paper\_dirs.py}}\fi}
\makeatother

\newcommand{\N}{\mathcal{N}}
\newcommand{\Nc}[1]{\mathcal{N}[#1]}
\newcommand{\Ui}{\mathcal{U}_i}
\newcommand{\Uj}{\mathcal{U}_j}
\newcommand{\Cc}{\mathcal{C}}
\newcommand{\W}{\mathcal W_1}
\newcommand{\s}{\mathsf{s}}
\newcommand{\tsink}{\mathsf{t}}
\newcommand{\Net}{\mathsf{G}}
\newcommand{\NetW}{\mathsf{G}^{\mathrm{w}}}
\newcommand{\NetT}{\widetilde{\Net}_3}
\newcommand{\kOR}{\kappa}
\newcommand{\Wit}{\Omega}
\newcommand{\li}{\ell_{ij}}
\newcommand{\dist}{\operatorname{dist}}
\DeclareMathOperator{\supp}{supp}
\newcommand{\methodname}{\texttt{orcut}}
\definecolor{Accent}{HTML}{1F77B4}
\definecolor{Relay}{HTML}{D62728}

\title{The Min-Cut Formulation of Ollivier--Ricci Curvature}

\author{Giorgio Micaletto\\
Department of Statistics, University of Chicago\\
\texttt{gmicaletto@uchicago.edu}}

\begin{document}

\maketitle

\begin{abstract}
Computing Ollivier--Ricci curvature normally requires a local distance matrix and a linear program per edge, prompting workarounds based on entropic regularization or combinatorial proxies. We show that exact curvature needs neither. For any one-step measures, the Kantorovich dual has an integral three-level optimum, reducing each edge to one minimum $\s$--$\tsink$ cut on a distance-free network linear in the read adjacency lists. This solves instances requiring quadratically many distance queries in linear time, while a converse reduction shows that no exact algorithm can beat bipartite matching. Our solver is a median \res{EThreePotSpeedupMedian} (up to \res{EThreePotSpeedupMax}) faster than distance construction plus \texttt{POT}'s C++ simplex, \res{EFiveGrcMixSlowdownMedian} faster than \texttt{GraphRicciCurvature}, and \res{EFourGraphflowOverFullMedian} faster than graph min-cost flow. The whole-graph times of \texttt{POT} are estimated from a degree-stratified sample of edges, with every edge whose local distance matrix would exceed $5\cdot10^7$ entries counted at zero time, so up to sampling error they are lower bounds. Even given the distances, Sinkhorn iterations at \texttt{GraphRicciCurvature}'s regularization are a median \res{EFiveSinkhornTenthSlowdownMedian} slower, and combinatorial proxies alter edge rankings, missing the bulk of the most negatively curved ones.
\end{abstract}

\ifdefined\orcutprovisional
\begin{center}\fcolorbox{red}{white}{\parbox{0.92\linewidth}{\footnotesize\color{red}\textbf{Provisional numbers.} Values marked $\dag$ come from a partial run of the experiment pipeline on toy graphs, not from the protocol of \cref{sec:experiments}.}}\end{center}
\fi

\section{Introduction}
\label{sec:intro}

Ollivier--Ricci (OR) curvature \citep{ollivier_ricci_2009,lin_curvature_2011} compares the Wasserstein-1 distance between random-walk measures at the two endpoints of an edge with the length of the edge, and it is used to study over-squashing in message-passing networks \citep{topping_squashing_2022,nguyen_revisiting_2023,fesser_mitigating_2024}, for structural encodings \citep{fesser_effective_2024}, for pooling \citep{sanders_pooling_2025}, to evaluate generative models \citep{southern_curvature_2023}, and to detect communities \citep{ni_community_2019,sia_ollivier_2019,tian_curvature_2025}.

The obstacle is cost. For $(i,j)\in E$ with degrees $d_i,d_j$, OR curvature requires a transport problem with up to $(d_i+1)(d_j+1)$ variables whose costs are graph distances. \texttt{GraphRicciCurvature} \citep{ni_community_2019} computes the distances by shortest paths and solves one linear program per edge with \texttt{POT} \citep{flamary_pot_2021}; for large neighbourhoods it switches to Sinkhorn iterations \citep{cuturi_sinkhorn_2013} and keeps at most $3{,}000$ neighbours per vertex. Thus, OR curvature is regarded as too expensive for large graphs or for algorithms that need to recompute it, and a growing literature replaces it by proxies \citep{forman_method_2003,sreejith_forman_2016,samal_comparative_2018,topping_squashing_2022,fesser_mitigating_2024,fesser_augmentations_2024} and approximations \citep{park_lower_2025,tian_curvature_2025,zhou_cycle_2026}, which trade exactness for speed, or bounds \citep{jost_clustering_2014,kang_transport_free_2026,gu_residual_2026,micaletto_scalable_2026}, which do not produce a point estimate. Moreover, \citet{dasgupta_computing_2023} prove a worst-case lower bound of $n^2/6$ expected distance-matrix queries for exact local algorithms, on neighbourhoods of size $n$, even on equal-degree ones.

We show that exact OR curvature needs neither the distance matrix nor a linear program. For measures supported on closed neighbourhoods, every relevant distance is decided by adjacency and common neighbours; the Kantorovich dual then has an integral three-level optimum that, written in unary, reduces to one minimum $\s$--$\tsink$ cut \citep{picard_maximal_1976,hochbaum_efficient_2001}, and on graph neighbourhoods this network is linear in the adjacency lists of $i$, $j$, and the vertices with net mass.

Integral potentials \citep[Lemma~2.2]{bhattacharya_exact_2020}, closure reductions for bounded difference constraints \citep{hochbaum_efficient_2001,ishikawa_exact_2003}, and $\{0,1,2,3\}$ local distances \citep{dasgupta_computing_2023} are known. Our contribution is a distance-free network (linear in the read adjacency lists via the graph's implicit biclique cover of distance-$2$ pairs \citep{feder_clique_1991}) and a bipartite-matching lower bound (\cref{thm:matching}): what remains is the cost of maximum flow.

\paragraph{Contributions.}
\begin{itemize}
\item[(C1)] \textbf{Reduction} (\cref{sec:reduction}). For costs in $\{1,\dots,D\}$, $\W=DA-F$ via a max-flow $F$ on $(D-1)(|P|+|N|)+2$ nodes (\cref{thm:layered}) with an exact certificate (\cref{prop:cert}) and a min--max formula (\cref{cor:hall}). One-step measures (lazy, non-lazy, any idleness, LLY \citep{lin_curvature_2011}, weighted walks) have $D=3$; $r$-step walks and $r$-balls have $D=2r+1$.
\item[(C2)] \textbf{Linear-size network} (\cref{sec:witness}). One relay per common neighbour bounds the network by $3\li+2$ nodes and $5\li$ arcs, built in $O(\li)$ time 
(\cref{thm:witness}, $\li=d_i+d_j+\sum_{x\in P\cup N}d_x$). For lazy, non-lazy, LLY, and fixed-idleness curvature, computations take $\li^{1+o(1)}$ time ($O(n)$ on the lower-bound instances of \citet{dasgupta_computing_2023}; \cref{prop:dasgupta}). Conversely, one edge encodes any bipartite maximum matching (\cref{thm:matching}), and the bound cannot be improved without a faster algorithm for it.
\item[(C3)] \textbf{Solver} (\cref{sec:solver}). A solver combining bitsets, block/witness relays, twin merging, warm-started Dinic with $O(\sqrt{A})$ phases (\cref{prop:phases}), and local incremental updates (\cref{lem:locality}).
\item[(C4)] \textbf{Evaluation} (\cref{sec:experiments}). Exactness on \res{EOneGraphs} graphs for the \res{EOneMeasures} measures of (C1), and speed on \res{EThreeGraphs} real networks of up to \res{EThreeLargestEdges} edges, against LP, network-simplex, min-cost flow, and approximate baselines.
\end{itemize}
Code: \url{https://github.com/GiorgioMB/orcut} (tag v1.4.4 produces every result here). The solver alone is the PyPI package \href{https://pypi.org/project/faster-or-curvature/}{\texttt{faster-or-curvature}}, which has its own version numbers: v1.2.2 contains the solver of \texttt{orcut} 1.4.4, 1.3.0 added the real lengths and weights of \cref{sec:conclusion}, and 1.3.1 is the finished release.

\section{Background and Related Work}
\label{sec:background}

\paragraph{Ollivier--Ricci curvature.}
Let $G=(V,E)$ be a finite, simple, connected, undirected graph with hop distance $\dist$, $\N(u)$ the neighbours of $u$, $\Nc{u}=\N(u)\cup\{u\}$, and $d_u=|\N(u)|$. 
Each $u\in V$ has a probability measure $m_u$, and the curvature of $(i,j)\in E$ is $\kOR(i,j)=1-\W(m_i,m_j)$. 
We call $m_u$ \emph{one-step} if $\supp m_u\subseteq\Nc{u}$; this covers $p$-lazy walks\footnote{This covers $m_u(u)=p$ and $m_u(x)=(1-p)/d_u$ for $x\in\N(u)$: the non-lazy walk at $p=0$, and the \emph{lazy walk}, the uniform measure on $\Nc{u}$, at $p=1/(d_u+1)$.} as well as weighted walks.
The LLY curvature $\kOR_{\rm LLY}=\lim_{p\to1}\kOR_p/(1-p)$ has $\kOR_p=(1-p)\,\kOR_{\rm LLY}$ for $p\ge 1/(\max\{d_i,d_j\}+1)$ \citep{bourne_idleness_2018}, and \texttt{GraphRicciCurvature}'s default is half the LLY curvature.

For $(i,j)\in E$ let $S=\Nc{i}\cup\Nc{j}$, $\Cc=\N(i)\cap\N(j)$, $\Ui=\N(i)\setminus\Nc{j}$, $\Uj=\N(j)\setminus\Nc{i}$, so that $S=\{i,j\}\sqcup\Cc\sqcup\Ui\sqcup\Uj$.  For $\mu=m_i-m_j$, let $P=\{\mu>0\}$, $N=\{\mu<0\}$, $\nu=-\mu$ on $N$, and $A=\mu(P)=\nu(N)$. Mass common to both measures can stay in place, so $\W(m_i,m_j)$ is the optimal cost of moving $\mu^+$ onto $\mu^-$ \citep{villani_topics_2003}, and by duality
\begin{equation}
\label{eq:dual}
\W(m_i,m_j)=\max\left\{\sum_{x\in P}\mu(x)u_x-\sum_{y\in N}\nu(y)v_y:u_x-v_y\le \dist(x,y)\ \ \forall (x,y)\in P\times N\right\}.
\end{equation}

\paragraph{Computing curvature.}
Transport solvers need the $|P|\times|N|$ cost matrix: network simplex \citep{bonneel_displacement_2011,flamary_pot_2021}, min-cost flow and assignment algorithms \citep{kuhn_hungarian_1955,ahuja_network_1993,perron_ortools_2024}, or entropic regularization \citep{cuturi_sinkhorn_2013} (which is approximate). The only formulation we know that avoids distances is Beckmann's: on a graph, $\W$ for the hop metric is the minimum cost of a flow on the edges with unit costs and divergence $\mu$ \citep{beckmann_continuous_1952}, \citep[\S~6.3]{peyre_computational_2019}; we use it as a baseline. Closed forms exist for special classes \citep{munch_ollivier_2019,bhattacharya_exact_2020,kelly_exact_2020,hehl_regular_2026,jung_combinatorial_2026}. 
\citet{dasgupta_computing_2023} observe that the local distances lie in $\{0,1,2,3\}$; for the lazy walk and equal degrees they cast the problem as a minimum-weight perfect matching with costs in $\{1,2,3\}$, reduce unequal degrees to such matchings, with an additive error when the smaller of $d_i+1$ and $d_j+1$ does not divide the larger (their Theorem~3), and bound the number of queries to the cost matrix that local algorithms need. 

\section{Transport on Small Integer Metrics Is a Minimum Cut}
\label{sec:reduction}

We state the reduction for a finite metric space $(X,\dist)$ and then specialize it. It is an instance of the classical reduction of integer programs with difference constraints and bounded variables to closures on layered graphs \citep{hochbaum_efficient_2001,ahuja_solving_2003,ishikawa_exact_2003}; what matters here is how effective it is on graph neighbourhoods. Let $\mu$ be a signed measure on $X$ with $\mu(X)=0$, and $\dist(x,y)\in\{1,\dots,D\}$ for all $(x,y)\in P\times N$.

\begin{definition}[Layered closure network]
\label{def:layered}
The \emph{layered closure network} $\Net_D(\mu)$ is the capacitated directed network with source $\s$, sink $\tsink$, and nodes $a_{x,\tau}$ ($x\in P$, $\tau\in\{2,\dots,D\}$) and $b_{y,\sigma}$ ($y\in N$, $\sigma\in\{1,\dots,D-1\}$). Its arcs are $\s\to a_{x,\tau}$ of capacity $\mu(x)$, $b_{y,\sigma}\to\tsink$ of capacity $\nu(y)$, and, of infinite capacity, the chains $a_{x,\tau+1}\to a_{x,\tau}$ ($\tau<D$) and $b_{y,\sigma+1}\to b_{y,\sigma}$ ($\sigma<D-1$) together with the implications $a_{x,\tau}\to b_{y,\tau-\dist(x,y)}$ for $(x,y)\in P\times N$ and $\tau\in\{\dist(x,y)+1,\dots,D\}$.
\end{definition}

\begin{theorem}
\label{thm:layered}
If $F$ is the value of a maximum $\s$--$\tsink$ flow in $\Net_D(\mu)$, then $\W(\mu^+,\mu^-)=D\,A-F$.
\end{theorem}

\begin{proof}[Proof sketch]
The constraint matrix of \cref{eq:dual} is totally unimodular, so \eqref{eq:dual} has an integral optimum (see \citealp[Lemma~2.2]{bhattacharya_exact_2020}), which can be normalized to $u\in\{1,\dots,D\}^P$ and $v\in\{0,\dots,D-1\}^N$. In unary, $u_x-v_y\le \dist(x,y)$ becomes the implications of \cref{def:layered}, so feasible potentials are the closed sets of the network, with weight $\mu(x)$ on the nodes $a_{x,\cdot}$ and $-\nu(y)$ on the nodes $b_{y,\cdot}$, and the objective is $A$ plus the weight of the closed set. A maximum-weight closed set weighs $(D-1)A-F$ \citep{picard_maximal_1976}. Details in \cref{app:proofs}.
\end{proof}

Only pairs at distance less than $D$ generate arcs, and no cost is stored: the network needs the relations ``$\dist(x,y)\le\tau$'' for $\tau<D$, which graph neighbourhoods make cheap.

\begin{lemma}
\label{lem:distances}
Let $(i,j)\in E$, $x\in\Nc{i}$, $y\in\Nc{j}$, $x\ne y$. Then $\dist(x,y)\le2$ unless $x\in\Ui$ and $y\in\Uj$, in which case $\dist(x,y)=1$ if $xy\in E$, $2$ if $xy\notin E$ and $\exists z:xz, yz\in E$, and $3$ otherwise.
\end{lemma}

Since $P\subseteq\Nc{i}$ and $N\subseteq\Nc{j}$ for one-step measures, \cref{lem:distances} implies $\dist(x,y)\in\{1,2,3\}$ on $P\times N$, so \cref{thm:layered} applies with $D=3$.

\begin{corollary}[One-step measures]
\label{cor:onestep}
For one-step measures $m_i,m_j$, $\W(m_i,m_j)=3A-F$, where $F$ is the maximum flow value of $\NetT$, obtained from $\Net_3(\mu)$ by dropping the arcs $b_{y,2}\to b_{y,1}$. If $\dist(x,y)\le 2$ on $P\times N$, then $\W(m_i,m_j)=2A-F_1$ for the maximum flow value $F_1$ of the bipartite network $\Net_2(\mu)$, and $\W(m_i,m_j)=2A$ if no pair is adjacent.
\end{corollary}

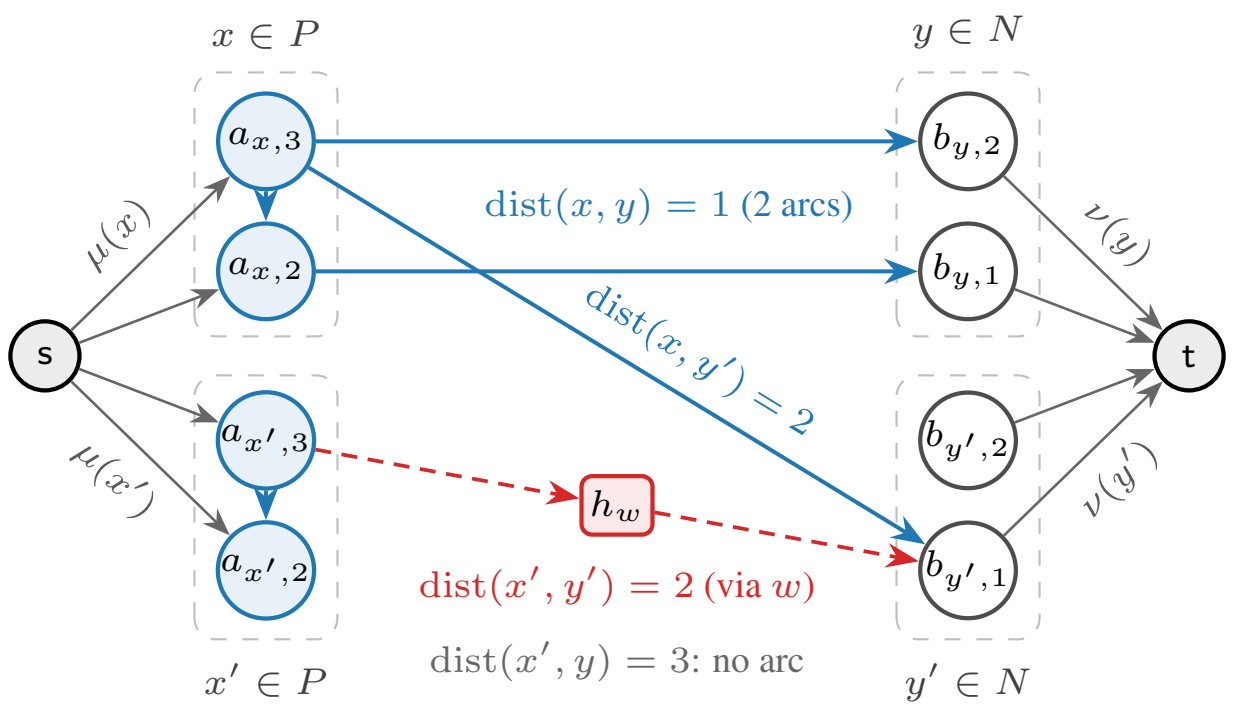
\begin{figure}[t]
\centering
\begin{tikzpicture}[>=Stealth, scale=0.86, every node/.style={font=\scriptsize},
  st/.style={circle,draw=black,thick,fill=gray!15,minimum size=13pt,inner sep=0pt},
  an/.style={circle,draw=Accent,thick,fill=Accent!10,minimum size=18pt,inner sep=0pt},
  bn/.style={circle,draw=black!70,thick,fill=white,minimum size=18pt,inner sep=0pt},
  hub/.style={rectangle,rounded corners=2pt,draw=Relay,thick,fill=Relay!10,minimum size=11pt,inner sep=1.5pt},
  grp/.style={draw=black!25,dashed,rounded corners=4pt},
  capa/.style={->,line width=0.5pt,draw=black!60},
  infa/.style={->,line width=0.75pt,draw=Accent},
  rela/.style={->,line width=0.7pt,draw=Relay,densely dashed}]
\draw[grp] (1.15, 0.15) rectangle (2.25, 2.18);
\draw[grp] (1.15,-2.18) rectangle (2.25,-0.15);
\draw[grp] (6.55, 0.15) rectangle (7.65, 2.18);
\draw[grp] (6.55,-2.18) rectangle (7.65,-0.15);
\node[above=1pt,text=black!75] at (1.7, 2.18) {$x\in P$};
\node[below=1pt,text=black!75] at (1.7,-2.18) {$x'\in P$};
\node[above=1pt,text=black!75] at (7.1, 2.18) {$y\in N$};
\node[below=1pt,text=black!75] at (7.1,-2.18) {$y'\in N$};

\node[st] (s) at (0,0) {$\s$};
\node[st] (t) at (8.8,0) {$\tsink$};
\node[an] (a13) at (1.7, 1.65) {$a_{x,3}$};
\node[an] (a12) at (1.7, 0.65) {$a_{x,2}$};
\node[an] (a23) at (1.7,-0.65) {$a_{x',3}$};
\node[an] (a22) at (1.7,-1.65) {$a_{x',2}$};
\node[bn] (b12) at (7.1, 1.65) {$b_{y,2}$};
\node[bn] (b11) at (7.1, 0.65) {$b_{y,1}$};
\node[bn] (b22) at (7.1,-0.65) {$b_{y',2}$};
\node[bn] (b21) at (7.1,-1.65) {$b_{y',1}$};
\node[hub] (h) at (4.4,-1.15) {$h_w$};

\draw[capa] (s) -- node[above,sloped,pos=0.45,text=black!65] {$\mu(x)$} (a13);
\draw[capa] (s) -- (a12);
\draw[capa] (s) -- (a23);
\draw[capa] (s) -- node[below,sloped,pos=0.45,text=black!65] {$\mu(x')$} (a22);
\draw[capa] (b12) -- node[above,sloped,pos=0.55,text=black!65] {$\nu(y)$} (t);
\draw[capa] (b11) -- (t);
\draw[capa] (b22) -- (t);
\draw[capa] (b21) -- node[below,sloped,pos=0.55,text=black!65] {$\nu(y')$} (t);

\draw[infa] (a13) -- (a12);
\draw[infa] (a23) -- (a22);
\draw[infa] (a13) -- (b12);
\draw[infa] (a12) -- (b11);
\node[text=Accent] at (4.8, 1.15) {$\dist(x,y)=1$ (2 arcs)};
\draw[infa] (a13) -- node[pos=0.60,sloped,above,text=Accent] {$\dist(x,y')=2$} (b21);

\draw[rela] (a23) -- (h);
\draw[rela] (h) -- (b21);
\node[below=2pt,text=Relay] at (h.south) {$\dist(x',y')=2$ (via $w$)};
\node[text=black!60] at (4.4,-2.35) {$\dist(x',y)=3$: no arc};
\end{tikzpicture}
\caption{$\NetT$ (\cref{cor:onestep}) for $x,x'\in P$ and $y,y'\in N$, grouped by vertex (dashed boxes). Grey arcs carry finite capacities ($\mu$ from $\s$, $\nu$ to $\tsink$, identical on both layers of a vertex); solid blue arcs have infinite capacity, encoding the vertical chains $a_{x,3}\to a_{x,2}$ and distance implications (two arcs for $\dist=1$, one for $\dist=2$, none for $\dist=3$). Dashed red: a relay $h_w$ (\cref{prop:relay}) replacing direct distance-$2$ arcs for all pairs in $P\times N$ sharing a common neighbour $w$ (\cref{thm:witness}).}
\label{fig:network}
\end{figure}

\noindent
The arc $b_{y,2}\to b_{y,1}$ is redundant: every arc into $b_{y,2}$ comes from an $a_{x,3}$ with $xy\in E$, which already implies $b_{y,1}$ via $a_{x,2}$ (\cref{app:proofs}). Thus $\NetT$ has $2(|P|+|N|)+2$ nodes and $3|P|+2|N|+2e_1+e_2$ arcs, where $e_1,e_2$ count pairs in $P\times N$ at distance $1,2$ (\cref{fig:network}). For $\supp m_u\subseteq B_r(u)$ (uniform balls, \citealp{ollivier_ricci_2009}; or $r$-step walks), all distances are at most $2r+1$ and $\Net_{2r+1}$ applies.

\begin{corollary}[A min--max formula]
\label{cor:hall}
For one-step measures and $X\subseteq P$, let $N_k(X)=\{y\in N:\dist(x,y)\le k\text{ for some }x\in X\}$. Then
\[
\W(m_i,m_j)=A+\max_{X_3\subseteq X_2\subseteq P}\Bigl[\mu(X_2)-\nu\bigl(N_1(X_2)\cup N_2(X_3)\bigr)+\mu(X_3)-\nu\bigl(N_1(X_3)\bigr)\Bigr].
\]
\end{corollary}

\noindent
Here $X_2,X_3$ are the vertices of potential at least $2$ and $3$. For equal degrees, matchings are known to determine curvature \citep[Section~2.2]{dasgupta_computing_2023}, \citep{hehl_regular_2026,jung_combinatorial_2026}; \cref{cor:hall} is the dual form for every one-step measure.

\begin{proposition}[Certificates]
\label{prop:cert}
Let $Z$ be the nodes reachable from $\s$ in the residual network of a maximum flow on $\Net_D(\mu)$. Then $u_x=1+|\{\tau:a_{x,\tau}\in Z\}|$ and $v_y=|\{\sigma:b_{y,\sigma}\in Z\}|$ solve \eqref{eq:dual}, and a maximum flow on the bipartite network with arcs $\s\to x$ ($\mu(x)$), $y\to\tsink$ ($\nu(y)$), and $x\to y$ ($\infty$) for $u_x-v_y=\dist(x,y)$ yields an optimal transport plan.
\end{proposition}

Checking a certificate needs the marginals of the plan, the distances on its support, and the constraints of \eqref{eq:dual}; for one-step measures only pairs with $u_x-v_y\ge2$ need checking, by adjacency and common-neighbour queries. Graph rewiring needs such plans \citep{nguyen_revisiting_2023}.

\section{A Network Linear in the Neighbourhood}
\label{sec:witness}

$\NetT$ still has one arc per pair at distance $2$, and $e_2$ can be of order $|P|\,|N|$; however, such pairs come in blocks: all pairs of vertices adjacent to a common vertex $w$ are within distance $2$, and one zero-weight node can carry their implications.

\begin{proposition}[Relays]
\label{prop:relay}
Let $P'\subseteq P$ and $N'\subseteq N$ with $\dist(x,y)\le2$ for all $(x,y)\in P'\times N'$. Adding a node $h$ of weight $0$ with infinite-capacity arcs $a_{x,3}\to h$ ($x\in P'$) and $h\to b_{y,1}$ ($y\in N'$), and deleting the arcs $a_{x,3}\to b_{y,1}$ for $(x,y)\in P'\times N'$, does not change the maximum flow value, also when relays are added one after another.
\end{proposition}

Call $w\in V$ (inside or outside $S$) a \emph{witness} of $(i,j)$ if it has neighbours in both $P$ and $N$. Let $\Wit$ be the set of witnesses and $\li=d_i+d_j+\sum_{x\in P\cup N}d_x$ the \emph{local input}, the number of adjacency entries read below.

\begin{definition}[Witness network]
\label{def:witness}
$\NetW$ has the nodes of $\NetT$ and a node $h_w$ per $w\in\Wit$, and the arcs of $\NetT$ except the arcs $a_{x,3}\to b_{y,1}$ of pairs at distance $2$, which are replaced by $a_{x,3}\to h_w$ for $x\in P\cap\N(w)$ and $h_w\to b_{y,1}$ for $y\in N\cap\N(w)$, with infinite capacity.
\end{definition}

\begin{theorem}
\label{thm:witness}
For one-step measures, (i) $\W(m_i,m_j)=3A-F^{\rm w}$, where $F^{\rm w}$ is the maximum flow value of $\NetW$; (ii) $\NetW$ has at most $3\li+2$ nodes and $5\li$ arcs; (iii) given the adjacency lists, $\NetW$ is built in $O(\li)$ time, with $O(|V|)$ workspace initialized once and reused for every edge.
\end{theorem}

By \cref{lem:distances}, every pair in $P\times N$ at distance $2$ shares a witness $w\in\Wit$, so applying \cref{prop:relay} to each $w\in\Wit$ turns $\NetT$ into $\NetW$ (\cref{app:proofs}).

\noindent
With integer masses (common denominator $\Lambda$, \cref{sec:solver}) and infinite capacities capped at $2A+1$, maximum flow on $\NetW$ takes $\li^{1+o(1)}\log\Lambda$ time, with high probability \citep{chen_maximum_2022} or deterministically \citep{vandenbrand_deterministic_2023}---hence $\li^{1+o(1)}$ whenever $\Lambda$ is polynomial in $\li$ (lazy, non-lazy, LLY, and fixed-idleness measures), without ever listing pairs at distance $2$ or $3$.

\paragraph{Graph min-cost flow.}
Beckmann's formulation on $ij$ and the edges incident to $P\cup N$ (\cref{lem:beckmann}) is already an almost-linear-time min-cost flow on at most $\li+1$ edges \citep{chen_maximum_2022}. \cref{thm:witness} instead eliminates edge costs, reducing $\W$ to a maximum flow that is faster in practice (\cref{sec:experiments}).

\paragraph{Relation to \citet{dasgupta_computing_2023}.}
Their lower bound of $n^2/6$ expected queries concerns queries to the complete bipartite cost graph $H=(L,R,w)$ of an equal-degree edge ($|L|=|R|=n$, $w=\dist\in\{1,2,3\}$). Their Proposition~2 realizes $H$ by a graph $G_H$ with $2n+1+e_1+2e_2$ edges ($e_1,e_2$ counting weight-$1$ and weight-$2$ pairs), with $e_2=0$ and $e_1\le n$ on their lower-bound instances.

\begin{proposition}
\label{prop:dasgupta}
(i) For $G_H$ and the lazy walk, $\NetW$ has $4n+e_2+2$ nodes and $5n+2e_1+2e_2$ arcs and is built in $O(n+e_1+e_2)$ time from adjacency lists; with unit masses, the minimum weight of a perfect matching of $H$ is $3n$ minus the maximum flow of $\NetW$.
(ii) On the instances of their lower-bound proofs, at most one phase of Dinic's algorithm finds the maximum flow, and exact curvature takes $O(n)$ time.
\end{proposition}

\noindent
The $\Omega(n^2)$ bound thus reflects the cost of querying a black-box distance matrix: from adjacency lists, the same instances take $O(n)$ time (\cref{tab:dasgupta}). Moreover, while (i) also follows for balanced matchings by decomposing weights in $\{1,2\}$ \citep{kao_decomposition_2001}, \cref{thm:layered,thm:witness} hold for unequal degrees and any one-step measure, whereas their matching reduction is exact only for the lazy walk when $\min\{d_i,d_j\}+1$ divides $\max\{d_i,d_j\}+1$.

\begin{theorem}[Curvature is at least as hard as bipartite matching]
\label{thm:matching}
Let $B=(L,R,E_B)$ be a bipartite graph with $|L|=|R|=n$, and let $G_B$ consist of the edge $(i,j)$, edges from $i$ to every vertex of $L$ and from $j$ to every vertex of $R$, and the edges $E_B$. For the lazy walk, $\W(m_i,m_j)=\bigl(3n-2\,\alpha'(B)\bigr)/(n+2)$, where $\alpha'(B)$ is the size of a maximum matching of $B$, and $\li=4n+2+2|E_B|$.
\end{theorem}

\noindent
An algorithm running in time $T(\li)$ thus finds $\alpha'(B)$ in time $T(O(n+|E_B|))+O(n+|E_B|)$, and the bounds $\li^{1+o(1)}$ (\cref{thm:witness}) and $O(\li\sqrt{|\Ui|})$ for the lazy walk with $d_i=d_j$ (\cref{prop:phases}) match those known for bipartite matching \citep{chen_maximum_2022,hopcroft_n52_1973}. No formula computable in linear time can equal the curvature on every edge unless bipartite matching takes linear time. $G_B$ is the graph $G_H$ of \citet[Proposition~2]{dasgupta_computing_2023} with weights in $\{1,3\}$.

\section{A Distance-Free Solver}
\label{sec:solver}

\Cref{thm:witness} bounds the network, but on most real neighbourhoods a smaller one is available. The solver (\cref{alg:solver}, \cref{app:impl}) combines both constructions.

\paragraph{Exact masses.}
Rows are the vertices of $\Nc{i}$ and columns those of $\Nc{j}$, with $d_i\ge d_j$. Masses are scaled to integers by a common denominator $\Lambda$, e.g.\ $(d_i+1)(d_j+1)$ for the lazy walk and $b\,d_id_j$ for idleness $a/b$ (\cref{app:impl}). Flows are integral and $\W=(3A-F)/\Lambda\in \mathbb Q$; edges with $\Lambda>2^{50}$ are skipped and flagged.

\paragraph{Bitsets and block relays.}
Columns are indexed by $\lceil (d_j+1)/64\rceil$ machine words. One pass over the lists of the columns gives, for every $w\in V$ adjacent to a column, the bitset $\mathrm{Col}(w)$ of its column neighbours. A row $x$ is adjacent to the columns $\mathrm{Col}(x)$ and within distance $2$ of $\mathrm{Cov}(x)=\bigcup_{w\in\N(x)}\mathrm{Col}(w)$, accumulated word by word and stopped once $\mathrm{Cov}(x)\supseteq\Uj$; by \cref{lem:distances} these give every distance. Most pairs at distance $2$ form blocks known in advance: every row is within distance $2$ of every column in $\{i,j\}\cup\Cc$, every row in $\{i,j\}\cup\Cc$ of every column, a \emph{full} row ($\mathrm{Cov}(x)\supseteq\Uj$) of all of $\Uj$, and a full column of all of $\Ui$. Four relays $H_1,\dots,H_4$ (\cref{prop:relay}) serve these blocks; only the remaining pairs need individual arcs.

\paragraph{Witness relays and bounded bitset work.}
When many remaining distance-$2$ pairs exist, they are routed through witness relays (\cref{def:witness}) whenever this shrinks the network by at least $4\times$ (checked via a fast upper bound). On large sparse neighbourhoods where bitsets cost $\lceil (d_j+1)/64\rceil$ words per adjacency entry but find few distance-$2$ pairs, we cap bitset work at $4$ words per row/column adjacency entry; if building or accumulating bitsets exceeds this while most rows are not full, the solver falls back directly to $\NetW$.

\paragraph{Twins, fast path, warm start, execution.}
Rows with the same adjacency to $N$, coverage, and block have the same successors and are merged, adding their capacities (\cref{prop:twins}); all leaves of a hub collapse into one row. If every row of $\Ui$ is full, no pair is at distance $3$ and one bipartite flow suffices (\cref{cor:onestep}). Otherwise a maximum flow of the one-layer adjacency network, routed along both layers, is a feasible flow of $\NetT$ from which the algorithm of \citet{dinic_algorithm_1970} starts. Edges run in parallel over threads with per-thread growable arenas and no allocation per edge.

\paragraph{Phases of Dinic's algorithm.}
Apart from the relays, every node carries at most twice the mass of its vertex, and no two relays are adjacent \citep{even_network_1975}, \citep[Theorem~4.2]{feder_clique_1991}; this bounds the phases.

\begin{proposition}[Phases]
\label{prop:phases}
Let $A$ be the total net mass in the smallest integer scaling. On $\NetT$, on $\NetW$, and on every network built by \cref{alg:solver}, Dinic's algorithm takes at most $2\sqrt{24A}+1$ phases and $O(\sqrt{A}\,m+A\log A)$ time, where $m$ is the number of arcs. For the lazy walk with $d_i=d_j$, $A=|\Ui|$, and computation takes $O(\li\sqrt{|\Ui|})$ time on $\NetW$.
\end{proposition}

\paragraph{Incremental updates.}
Many algorithms change only a few edges per iteration (e.g.\ rewiring).

\begin{lemma}[Locality]
\label{lem:locality}
Let $G'$ be obtained from $G$ by inserting or deleting vertex pairs $(a_l,b_l)$, $l=1,\dots,k$, and let the measure of each vertex depend only on its neighbourhood (all of the above do, with fixed weights on unchanged edges for weighted walks). The curvature of an edge $(i,j)$ of $G'$ can differ from its value in $G$ only if $i$ or $j$ lies in $U=\bigcup_{l}(\Nc{a_l}\cup\Nc{b_l})$, taken in $G'$.
\end{lemma}

\section{Experiments}
\label{sec:experiments}
We evaluate exactness (E1), per-edge scaling (E2), whole-graph speed (E3), ablations (E4), approximation errors (E5), and additional measures (E6, \cref{app:more}). Unless noted, we use the lazy walk with \res{EThreeThreads} threads on one node of a Slurm cluster (\cref{app:setup}).

\paragraph{Baselines and data.} \texttt{POT}'s network simplex \citep{flamary_pot_2021} (C++ and \texttt{ot.emd2}); \texttt{OR-Tools}' integer min-cost flow \citep{perron_ortools_2024}; \texttt{HiGHS} \citep{huangfu_parallelizing_2018}; min-cost flow on the graph itself (\cref{lem:beckmann}; \texttt{OR-Tools}); default \texttt{GraphRicciCurvature} 0.6.1 (GRC; \citealp{ni_community_2019}); entropic OT \citep{cuturi_sinkhorn_2013}. Transport solvers are \emph{given} the local distances, built by our fastest code and timed separately. On whole graphs, the time of \texttt{POT}, distance construction included, is estimated from a sample of edges stratified by $\lfloor\log_2(d_i+d_j)\rfloor$. Edges whose local distance matrix would have $(d_i+1)(d_j+1)>5\cdot10^7$ entries are left out and count at zero time. From each stratum, up to $50$ of the remaining edges are drawn uniformly at random and timed once each, single-threaded and with no time limit. Each stratum is charged the number of its remaining edges times the mean time of its sample, and the sum over strata is divided by the number of threads. Counting the left-out edges at zero and dividing by the number of threads can only lower this estimate, so up to sampling error it is a lower bound on the parallel time of \texttt{POT}. 

Data: \res{EThreeGraphs} real networks and seeded random graphs (\cref{app:setup}).
The proxies are BF, the Balanced Forman curvature of \citet{topping_squashing_2022}; $\mathrm{AF}_3=4-d_i-d_j+3|\Cc|$, the triangle-augmented Forman curvature of \citet{fesser_mitigating_2024}; and the \citet[][henceforth JL]{jost_clustering_2014} mean (the midpoint of the JL bounds on non-lazy curvature), the surrogate of \citet{tian_curvature_2025}.

\subsection{Exactness (E1)}
On \res{EOneGraphs} graphs and \res{EOneMeasures} measures (lazy, non-lazy, LLY, idleness $1/2$, a weighted lazy walk), \cref{alg:solver} agrees with \texttt{OR-Tools}' integer min-cost flow on all \res{EOneOrtoolsChecked} edges checked, with \texttt{POT} up to \res{EOnePotMaxDiff}, and (for the lazy, non-lazy, and LLY curvatures) with the oracle of \cref{app:impl}, and the layered closure of \cref{thm:layered} on every edge. All \res{EOneCertChecked} certificates were verified (\cref{tab:exactness}). The witness network and every configuration of E4 return the same values.

\subsection{Cost of one edge (E2)}

\begin{figure}
\centering
\includegraphics[width=\textwidth]{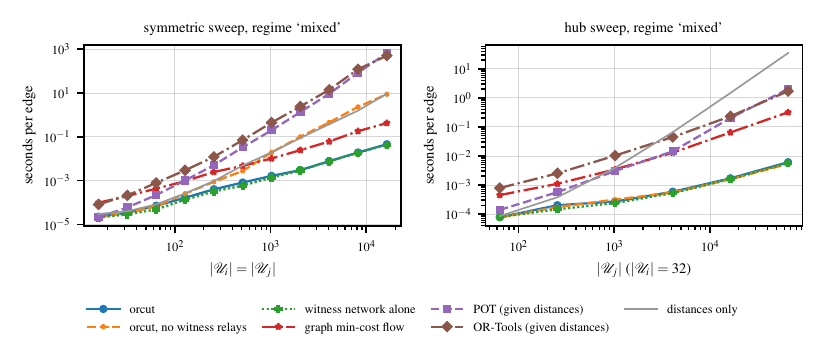}
\caption{Time per edge on a neighbourhood where half of $\Ui\times\Uj$ is at distance $2$ (see also \cref{fig:per-edge-all}). Transport solvers are given the distances (grey: their construction); other times include everything.}
\label{fig:per-edge}
\end{figure}

\Cref{fig:per-edge} isolates one edge with prescribed $|\Ui|$, $|\Uj|$, common neighbours, and distance structure (five regimes, \cref{app:setup}). At $|\Ui|=|\Uj|=\res{ETwoSymMaxS}$ the solver is \res{ETwoSymPotMin}--\res{ETwoSymPotMax} faster than \texttt{POT} given the distances (up to \res{ETwoSymPotBuildMax} counting their construction), up to \res{ETwoSymOrtoolsMax} faster than \texttt{OR-Tools}, and \res{ETwoSymGraphflowMin}--\res{ETwoSymGraphflowMax} faster than graph min-cost flow; on ``hub'' edges it is up to \res{ETwoHubPotMax} faster than \texttt{POT} given the distances. 
One flag switches off the witness relays, the cap on bitset work, and the fallback to $\NetW$; without them the solver is \res{ETwoSymNoWitnessMin}--\res{ETwoSymNoWitnessMax} slower at that size (\cref{tab:witness}): most in regime ``mixed'', where many pairs are at distance 2 and few rows are full, and also in regime ``tree'', where no pair is at distance 2 and the cost is the uncapped bitset work. In the dense regime, which uses none of them, the difference is noise.
On all controlled edges the witness network has at most \res{ETwoWitnessArcsPerInputMax} arcs per unit of local input (\cref{thm:witness}: $5$) and grows with exponent \res{ETwoWitnessArcSlopeMax} in $\li^S=\sum_{x\in S}d_x$, against up to \res{ETwoNoWitnessArcSlopeMax} with explicit pairs (\cref{tab:witness}). On real graphs, weighting degree strata by their share of edges, the median gain over \res{ETwoRealGraphs} graphs is \res{ETwoRealPotMedian} against \texttt{POT} given the distances (\res{ETwoRealPotBuildMedian} counting their construction) and \res{ETwoRealGraphflowMedian} against graph min-cost flow.

\begin{table}
\caption{Lower-bound instances of \citet{dasgupta_computing_2023} and related families (\cref{app:setup}): local input, arcs, and times at the largest $n$, and times of our solver and of \texttt{POT} with its distances at the largest $n$ whose $n^2$ distances fit the memory guard; $k$: fitted exponent of our time $\propto n^k$. Arcs: those of the network the solver builds, after twin merging and the other reductions of \cref{app:impl}; \cref{prop:dasgupta}(i) counts $\NetW$ before them.}
\label{tab:dasgupta}
\centering
\small
\setlength{\tabcolsep}{4pt}
\begin{tabular}{@{}>{\raggedright\arraybackslash\hangindent=0.6em\hangafter=1}p{2.3cm}rrrrrrrr@{}}
\toprule
& & & & \multicolumn{2}{c}{largest $n$} & \multicolumn{2}{c}{$n=\res{EDasPotMaxN}$} & \\
\cmidrule(lr){5-6}\cmidrule(lr){7-8}
Family & $n$ & $\li$ & arcs & \methodname & graph flow & \methodname & \shortstack[r]{\texttt{POT} +\\distances} & $k$ \\
\midrule
\resrows{dasgupta}
\bottomrule
\end{tabular}
\end{table}

\paragraph{Lower-bound instances and beyond.}
\Cref{tab:dasgupta} runs four families up to $n=\res{EDasMaxN}$: one pair and a perfect matching at distance $1$ \citep[Theorem~4]{dasgupta_computing_2023}, graphs $G_H$ with bounded weighted degrees (the class of their approximation algorithms, their Theorem~5), and, beyond $G_H$, $\sqrt n$ shared middle vertices that put $n^{3/2}$ pairs at distance $2$ with $O(n)$ edges. Our time grows as $n^{k}$ with $k$ between \res{EDasSlopeMin} and \res{EDasSlopeMax}, and the slowest family takes \res{EDasOursMaxTime} at the largest $n$; \texttt{POT} is \res{EDasPotOverOursMin}--\res{EDasPotOverOursMax} slower already at $n=\res{EDasPotMaxN}$, beyond which the distances exceed the memory guard, and graph min-cost flow is \res{EDasGraphflowOverOursMin}--\res{EDasGraphflowOverOursMax} slower.

\subsection{Whole graphs (E3) and ablation (E4)}

\begin{table}[t]
\caption{Lazy curvature of all edges (\res{EThreeThreads} threads). \texttt{POT} (est.): lower-bound estimate for distances plus C++ network simplex, with edges whose local distance matrix would exceed $5\cdot10^7$ entries at zero time (\cref{sec:experiments}). Memory: resident-set growth (MB, sampled every 50\,ms) during a timed run after warm-up, excluding reused memory ($0$: no growth beyond one page). Twins: rows per merged row on edges with an endpoint of degree $\ge100$.}
\label{tab:large}
\centering
\small
\setlength{\tabcolsep}{4pt}
\begin{tabular}{@{}lrrrrrrrr@{}}
\toprule
Graph & $n$ & $m$ & $d_{\max}$ & \methodname & \texttt{POT} (est.) & speedup & memory & twins \\
\midrule
\resrows{large}
\bottomrule
\end{tabular}
\end{table}

\Cref{tab:large} reports every edge of \res{EThreeGraphs} real networks; the largest, \res{EThreeLargestName} with \res{EThreeLargestEdges} edges, takes \res{EThreeLargestTime}. Against the lower-bound estimate for distances plus \texttt{POT}, the speedup is \res{EThreePotSpeedupMin}--\res{EThreePotSpeedupMax} (median \res{EThreePotSpeedupMedian}), largest on graphs with hubs, where twin merging compresses up to \res{EThreeTwinMax} rows into one, and smallest on dense, clustered neighbourhoods. A median \res{EThreeShareFastMedian} of the edges take the one-layer fast path and at most \res{EThreeShareWitnessMax} use witness relays. On the \res{EThreeOtherMeasuresGraphs} networks of \cref{tab:large} other than Orkut, the non-lazy and LLY curvatures cost at most \res{EThreeOtherMeasuresRelMax} the lazy one, and on the graphs of E6 at most \res{ESixNonLazyLLYRelMax} (\cref{tab:measures}); the solver scales by \res{EThreeScalingMin}--\res{EThreeScalingMax} on \res{EThreeScalingThreads} threads (\res{EThreeScalingDBLP} on DBLP, \res{EThreeScalingGitHub} on GitHub, \res{EThreeScalingYouTube} on YouTube, \res{EThreeScalingLiveJournal} on LiveJournal). On random graphs (\cref{fig:synth,tab:synth}), the time grows about linearly with $m$ at fixed average degree, and faster where hubs grow with $n$ (\citeauthor{barabasi_emergence_1999} and \citeauthor{chung_random_2002} graphs), as the local input of each edge (\cref{thm:witness}) grows with the degrees.

\begin{table}
\caption{Single-threaded time relative to the full solver, extrapolated from a stratified sample of edges (graphs of E4 in \cref{app:setup}; oracle of \cref{app:impl}). --: not estimated (the oracle ran out of time on GitHub; \texttt{POT}'s local problems exceed the memory guard on CL2.1 and GitHub).}
\label{tab:ablation}
\centering
\footnotesize
\setlength{\tabcolsep}{3pt}
\begin{tabular}{@{}l*{10}{c}@{}}
\toprule
\resrows{ablation-head}
\midrule
\resrows{ablation}
\bottomrule
\end{tabular}
\end{table}

\Cref{tab:ablation} switches the steps of \cref{sec:solver} on and off. The two-layer cut on bitsets alone is a median \res{EFourCutOverFullMedian} slower than the full solver (up to \res{EFourCutOverFullMax}); relays matter on dense neighbourhoods, twin merging on graphs with hubs (up to \res{EFourTwinsMax}), and the warm start up to \res{EFourNoWarmOverFullMax}. Witness relays change little on these graphs (\res{EFourNoWitnessOverFullMin}--\res{EFourNoWitnessOverFullMax}): real neighbourhoods are small or clustered and the bound on the bitset work is not reached, so they serve the large sparse neighbourhoods of E2. Linear size is not the whole story: the witness network alone is a median \res{EFourWitnessAloneOverFullMedian} slower than the full solver, graph min-cost flow \res{EFourGraphflowOverFullMin}--\res{EFourGraphflowOverFullMax} (median \res{EFourGraphflowOverFullMedian}), and distances plus \texttt{POT} \res{EFourPotOverFullMin}--\res{EFourPotOverFullMax} on the \res{EFourPotGraphs} graphs that fit in memory.

\subsection{Approximate pipelines (E5)}

\begin{table}
\caption{Approximations against exact curvature over \res{EFiveGraphs} graphs: GRC methods (\cref{app:setup}); proxies by rank against the LLY curvature, and the JL mean also by error against the non-lazy one, which it estimates. MAE: median over graphs; max: worst edge; sign: share of edges whose sign differs; $\rho$: Spearman; rec.: share of the $1\%$ most negatively curved edges recovered.}
\label{tab:approx}
\centering
\small
\begin{tabular}{@{}lccccc@{}}
\toprule
Method & MAE & max & sign & $\rho$ & rec. \\
\midrule
\resrows{approx}
\bottomrule
\end{tabular}
\end{table}

GRC with its defaults is exact up to floating point on most edges, but it keeps only $3{,}000$ neighbours of a vertex and switches to Sinkhorn when both endpoints have degree above $2{,}000$; on graphs with such hubs its error reaches \res{EFiveGrcMixHubMaxErr}, and it is a median \res{EFiveGrcMixSlowdownMedian} slower than our solver with the same threads. 
On the local problems themselves, at GRC's idleness $1/2$ (\cref{fig:tradeoff}), entropic OT at $\varepsilon=0.1$, GRC's choice\footnote{GRC's own Sinkhorn, run on every edge with GRC's truncation to $3{,}000$ neighbours, has the larger MAE of \cref{tab:approx}, also at idleness $1/2$.}, has a median MAE of \res{EFiveSinkhornTenthMaeMedian} at \res{EFiveSinkhornTenthSlowdownMedian} our time, and at $\varepsilon=0.01$ a negligible error at \res{EFiveSinkhornSmallSlowdownMedian}.

Proxies are cheap but rank edges differently from the curvature they stand for (\cref{tab:approx}). Of the $1\%$ most negatively curved edges, BF, $\mathrm{AF}_3$, and the JL mean recover only \res{EFiveRecallBF}, \res{EFiveRecallAFThree}, and \res{EFiveRecallJL}.
We do not measure the accuracy of downstream methods, like graph rewiring, as those gains are contested \citep{tori_effectiveness_2025}; what the approximations change is which edges they act on.

\section{Limitations and Conclusion}
\label{sec:conclusion}

The reduction needs integer distances on a small range. \Cref{thm:layered} applies to integer edge lengths, with $D$ the largest local distance; real-valued lengths, as in Ricci flows \citep{ni_community_2019}, are not covered. 
Version 1.3.1 of the PyPI package \texttt{faster-or-curvature} implements real edge lengths, through a network simplex on the local problem, and real walk weights, through the same cut with real capacities, both in floating point.
We leave it out of this paper, as it adds engineering rather than a new formulation; in the benchmarks of the package it is $2.6$ to $23$ times faster than \texttt{GraphRicciCurvature} 0.6.1 on weighted networks, with the same values to $10^{-13}$.
The local input can exceed $|P|\,|N|$ when the neighbours of $i$ and $j$ have large degrees, and our maximum flows use Dinic's algorithm, whose running time grows with the square root of the total mass (\cref{prop:phases}) rather than being almost linear. For $r\ge2$, our implementation still builds distances. Within these limits, OR curvature need not be the expensive part of curvature-based graph learning: one minimum cut per edge gives the value, a certificate, and an update rule after edge changes, while the approximations used to avoid it were, in our settings, either slower or ranked the edges differently, missing many of the most negatively curved edges on which curvature-based algorithms act.

\section*{Acknowledgements}
We acknowledge the computational resources provided by the Aalto Science-IT project.

{
\small
\bibliography{reference}
\bibliographystyle{plainnat}
}

\appendix
\sloppy

\section{Proofs}
\label{app:proofs}

Throughout, $\mu$ is a signed measure on a finite metric space $(X,\dist)$ with $\mu(X)=0$, $P=\{\mu>0\}$, $N=\{\mu<0\}$, $\nu=-\mu|_N$, $A=\mu(P)=\nu(N)>0$ (if $A=0$ every statement is trivial), and $\dist(x,y)\in\{1,\dots,D\}$ on $P\times N$.

\subsection{\texorpdfstring{\Cref{thm:layered}}{Theorem 1}}

\begin{proof}
\emph{Duality.} $\W(\mu^+,\mu^-)$ is the value of the transportation problem $\min\{\sum_{x,y}\dist(x,y)\pi_{xy}:\pi\ge0,\ \sum_y\pi_{xy}=\mu(x),\ \sum_x\pi_{xy}=\nu(y)\}$, which is feasible and bounded; its dual is \eqref{eq:dual}, with the same value. The constraint matrix of \eqref{eq:dual} is the transpose of the incidence matrix of the complete bipartite graph on $P\sqcup N$, which is totally unimodular, and the right-hand sides $\dist(x,y)$ are integers, so \eqref{eq:dual} has an integral optimal solution $(u,v)$ \citep{ahuja_network_1993}.

\emph{Range.} Let $v_0=\min_{y\in N}v_y$ and $\bar u_x=\min_{y\in N}\{v_y+\dist(x,y)\}$. Then $\bar u_x\ge u_x$ by feasibility, $(\bar u,v)$ is feasible by construction, and its value is not smaller because $\mu(x)>0$; moreover $v_0+1\le\bar u_x\le v_0+D$, since $\dist(x,y)\ge1$ and $v_y\ge v_0$, and $\dist(x,y)\le D$ at a minimizer of $v$. Let $\bar v_y=\min\{v_y,\,v_0+D-1\}$. If $\bar v_y=v_y$, the constraints of $y$ are unchanged; otherwise $\bar u_x-\bar v_y\le(v_0+D)-(v_0+D-1)=1\le \dist(x,y)$. Since $\bar v_y\le v_y$ and $\nu(y)>0$, the value does not decrease. Finally, subtracting $v_0$ from all potentials preserves feasibility and the value, because $\mu(P)-\nu(N)=0$. Hence \eqref{eq:dual} has an optimal solution with $u\in\{1,\dots,D\}^P$ and $v\in\{0,\dots,D-1\}^N$, and we may restrict \eqref{eq:dual} to such vectors.

\emph{Unary encoding.} For such $(u,v)$ let $\alpha_{x,\tau}=[u_x\ge\tau]$ for $\tau\in\{2,\dots,D\}$ and $\beta_{y,\sigma}=[v_y\ge\sigma]$ for $\sigma\in\{1,\dots,D-1\}$. This is a bijection onto the 0/1 vectors that are non-increasing in $\tau$ and in $\sigma$, with $u_x=1+\sum_\tau\alpha_{x,\tau}$ and $v_y=\sum_\sigma\beta_{y,\sigma}$. For $(x,y)\in P\times N$ with $d=\dist(x,y)$, we claim that $u_x-v_y\le d$ if and only if $\alpha_{x,\tau}\le\beta_{y,\tau-d}$ for all $\tau\in\{d+1,\dots,D\}$. If $u_x-v_y\le d$ and $\alpha_{x,\tau}=1$, then $v_y\ge u_x-d\ge\tau-d$, so $\beta_{y,\tau-d}=1$. Conversely, if $u_x\le d$ the constraint holds since $v_y\ge0$; otherwise $\tau=u_x\in\{d+1,\dots,D\}$ has $\alpha_{x,\tau}=1$, so $\beta_{y,u_x-d}=1$, that is, $v_y\ge u_x-d$.

\emph{Closure.} Give the node $a_{x,\tau}$ the weight $\mu(x)>0$ and the node $b_{y,\sigma}$ the weight $-\nu(y)<0$. By the previous step, feasible pairs $(u,v)$ in the restricted range correspond to the sets $Z$ of nodes (those whose indicator is one) that are closed under the arcs of \cref{def:layered} other than those incident to $\s$ and $\tsink$: the chains encode monotonicity and the implication arcs the constraints. The objective of \eqref{eq:dual} equals $\sum_x\mu(x)(1+\sum_\tau\alpha_{x,\tau})-\sum_y\nu(y)\sum_\sigma\beta_{y,\sigma}=A+w(Z)$. By \citet{picard_maximal_1976}, the maximum weight of a closed set equals the sum of the positive weights minus the capacity of a minimum cut in the network with arcs $\s\to z$ of capacity $w(z)$ for $w(z)>0$, $z\to\tsink$ of capacity $-w(z)$ for $w(z)<0$, and the closure arcs with infinite capacity, which is $\Net_D(\mu)$. The positive weights sum to $(D-1)A$, so $\W=A+(D-1)A-F=DA-F$ by the max-flow min-cut theorem.
\end{proof}

\noindent
\cref{thm:layered} holds for any problem with costs in $\{1,\dots,D\}$, which we use for $H$ in \cref{prop:dasgupta}.

\subsection{\texorpdfstring{\Cref{lem:distances}, \cref{cor:onestep}, and \cref{cor:hall}}{Lemma 2 and Corollaries 3 and 4}}

\begin{proof}[Proof of \cref{lem:distances}]
If $y\in\Nc{i}$, then $x,y\in\Nc{i}$ and $\dist(x,y)\le2$ through $i$. If $x\in\Nc{j}$, then $\dist(x,y)\le2$ through $j$.
Otherwise, $x\in\Nc{i}\setminus\Nc{j}=\Ui$ and $y\in\Nc{j}\setminus\Nc{i}=\Uj$. The path $x,i,j,y$ gives $\dist(x,y)\le3$; moreover, $\dist(x,y)=1$ if and only if $xy\in E$, and $\dist(x,y)\le2$ if and only if $xy\in E$ or $x$ and $y$ have a common neighbour.
\end{proof}

\begin{proof}[Proof of \cref{cor:onestep}]
If $\mu(x)>0$ then $m_i(x)>0$, so $P\subseteq\Nc{i}$; likewise $N\subseteq\Nc{j}$, and $P\cap N=\emptyset$. By \cref{lem:distances}, $\dist(x,y)\in\{1,2,3\}$ on $P\times N$, so \cref{thm:layered} applies with $D=3$; its implication arcs are $a_{x,2}\to b_{y,1}$ and $a_{x,3}\to b_{y,2}$ for $\dist(x,y)=1$, and $a_{x,3}\to b_{y,1}$ for $\dist(x,y)=2$. It remains to show that the chain arcs $b_{y,2}\to b_{y,1}$ can be dropped. Every set closed in $\Net_3(\mu)$ is closed without them. Conversely, let $Z$ be a maximum-weight closed set of the network without these arcs, and suppose $b_{y,2}\in Z$. If no $a_{x,3}$ with $xy\in E$ were in $Z$, no arc would enter $b_{y,2}$ from $Z$, and $Z\setminus\{b_{y,2}\}$ would be closed and heavier (the weight of $b_{y,2}$ is negative), a contradiction. So $a_{x,3}\in Z$ for some $x$ adjacent to $y$, hence $a_{x,2}\in Z$ and $b_{y,1}\in Z$. Thus maximum-weight closed sets of the two networks coincide in weight, and so do the minimum cuts. If no pair is at distance $3$, then $\dist(x,y)\in\{1,2\}$ on $P\times N$ and \cref{thm:layered} with $D=2$ gives $\W=2A-F_1$, where $\Net_2$ has one node per vertex and the arcs $a_{x,2}\to b_{y,1}$ for $\dist(x,y)=1$; with no adjacent pair, $F_1=0$.
\end{proof}

\begin{proof}[Proof of \cref{cor:hall}]
By the proofs of \cref{thm:layered} and \cref{cor:onestep}, $\W=A+w(Z)$ for a maximum-weight set $Z$ closed in $\NetT$. Let $X_3=\{x:a_{x,3}\in Z\}$ and $X_2=\{x:a_{x,2}\in Z\}$; the chains give $X_3\subseteq X_2$. Closedness requires $b_{y,1}\in Z$ for $y\in N_1(X_2)$ (arcs $a_{x,2}\to b_{y,1}$) and for $y\in N_2(X_3)$ (arcs $a_{x,3}\to b_{y,1}$ at distance $2$, and $a_{x,3}\to a_{x,2}\to b_{y,1}$ at distance $1$), and $b_{y,2}\in Z$ for $y\in N_1(X_3)$; these sets of $b$-nodes are closed, and since $b$-nodes have negative weight, a maximum-weight $Z$ contains no other. Its weight is the bracket.
\end{proof}

\noindent\emph{Balls and multi-step walks.} If $\supp m_i\subseteq B_r(i)$ and $\supp m_j\subseteq B_r(j)$, then $P\subseteq B_r(i)$, $N\subseteq B_r(j)$, and $\dist(x,y)\le r+1+r$ for $(x,y)\in P\times N$, so \cref{thm:layered} applies with $D=2r+1$.

\subsection{\texorpdfstring{\Cref{prop:cert}}{Proposition 5}}

\begin{proof}
Let $f$ be a maximum flow. The set $Z$ of nodes reachable from $\s$ in the residual network is the source side of a minimum cut; it contains no tail of an infinite arc whose head is outside $Z$ (the cut would be infinite), so $Z\setminus\{\s\}$ is closed and, by the proof of \cref{thm:layered} and \citet{picard_maximal_1976}, of maximum weight. Since the chains are arcs of $\Net_D$, the sets $\{\tau:a_{x,\tau}\in Z\}$ and $\{\sigma:b_{y,\sigma}\in Z\}$ are initial segments, and the decoding $u_x=1+|\{\tau:a_{x,\tau}\in Z\}|$, $v_y=|\{\sigma:b_{y,\sigma}\in Z\}|$ is the inverse of the unary encoding; hence $(u,v)$ is feasible and optimal for \eqref{eq:dual}. By complementary slackness, every optimal transport plan is supported on the tight set $T=\{(x,y):u_x-v_y=\dist(x,y)\}$, and an optimal plan exists, so the transportation problem restricted to $T$ is feasible, and a maximum flow of the bipartite network of the statement saturates the arcs out of $\s$, i.e.\ has value $A$; its arc flows form a plan $\pi$ with the right marginals and support in $T$, whose cost is $\sum\pi_{xy}(u_x-v_y)=\sum\mu(x)u_x-\sum\nu(y)v_y$, the optimal dual value. Conversely, any plan and any feasible potential with equal values prove that this number is $\W$ (weak duality).
\end{proof}

\subsection{\texorpdfstring{\Cref{prop:relay}}{Proposition 6} and twin merging}

\begin{proof}[Proof of \cref{prop:relay}]
Call a network \emph{admissible} if it consists of the nodes and the source and sink arcs of $\NetT$, possibly extra nodes of weight $0$, and infinite-capacity arcs, such that the restrictions of its closed sets to the nodes of $\NetT$ are exactly the closed sets of $\NetT$. Admissible networks have the maximum closure weight of $\NetT$, hence its maximum flow value (closed sets correspond to cuts as in the proof of \cref{thm:layered}; the extra nodes have weight $0$). $\NetT$ is admissible; we show that adding one relay to an admissible network $\Net$ gives an admissible network $\Net'$.
Let $Z'$ be closed in $\Net'$. Then $Z'\setminus\{h\}$ is closed in $\Net$: every arc of $\Net$ is an arc of $\Net'$ except the deleted arcs $a_{x,3}\to b_{y,1}$, $(x,y)\in P'\times N'$, and for those $a_{x,3}\in Z'$ implies $h\in Z'$ and then $b_{y,1}\in Z'$. Conversely, let $Z_0$ be closed in $\NetT$ and $Z$ a closed set of $\Net$ restricting to $Z_0$. If $Z$ contains no $a_{x,3}$ with $x\in P'$, it is closed in $\Net'$. Otherwise $Z\cup\{h\}$ is closed in $\Net'$, because $b_{y,1}\in Z_0$ for every $y\in N'$: $\NetT$ has the arc $a_{x,3}\to b_{y,1}$ if $\dist(x,y)=2$ and the path $a_{x,3}\to a_{x,2}\to b_{y,1}$ if $\dist(x,y)=1$. Both directions preserve the restriction to the nodes of $\NetT$, so $\Net'$ is admissible.
\end{proof}

The four relays of \cref{sec:solver} satisfy the hypothesis by \cref{lem:distances}: $H_1$ joins all rows (in $\Nc{i}$) to the columns in $\{i,j\}\cup\Cc\subseteq\Nc{i}$; $H_2$ joins the rows in $\{i,j\}\cup\Cc\subseteq\Nc{j}$ to all columns (in $\Nc{j}$); $H_3$ joins the full rows to $\Uj$; and $H_4$ joins the remaining rows of $\Ui$ to the full columns. Together with the direct arcs, or the witness relays, for the remaining pairs at distance $2$, they implement every implication $a_{x,3}\Rightarrow b_{y,1}$ of a pair within distance $2$, and no other.

\begin{proposition}[Twins]
\label{prop:twins}
Let $x,x'\in P$ be such that, for each $\tau$, the nodes $a_{x,\tau}$ and $a_{x',\tau}$ have the same successors apart from their chain successors $a_{x,\tau-1}$ and $a_{x',\tau-1}$. Replacing, for each $\tau$, the two nodes by one node with source capacity $\mu(x)+\mu(x')$ does not change the maximum flow value.
\end{proposition}

\begin{proof}
Let $Z$ be a maximum-weight closed set. If $a_{x,\tau}\in Z$, then adding $a_{x',\sigma}$ for all $\sigma\le\tau$ keeps $Z$ closed (their successors are those of $a_{x,\sigma}$, which are in $Z$ by the chain) and does not decrease the weight, which is positive on these nodes; by symmetry, there is a maximum-weight closed set that contains, for every $\tau$, both $a_{x,\tau}$ and $a_{x',\tau}$ or neither. Such sets are exactly the closed sets of the merged network, with the same weights.
\end{proof}

In the solver, the successors of a row are determined by its adjacency to $N$, its coverage $\mathrm{Cov}$ (which fixes whether it is full and its arcs at distance $2$), and whether it lies in $\{i,j\}\cup\Cc$ (which fixes its relays), so rows with equal keys satisfy the hypothesis; in the witness network, by its adjacent columns and its witnesses.

\subsection{\texorpdfstring{\Cref{thm:witness}}{Theorem 7} and graph min-cost flow}

\begin{proof}[Proof of \cref{thm:witness}]
(i) Let $\Wit=\{w_1,\dots,w_k\}$ and $P_l=P\cap\N(w_l)$, $N_l=N\cap\N(w_l)$. For $(x,y)\in P_l\times N_l$ we have $x\ne y$ (as $P\cap N=\emptyset$) and the path $x,w_l,y$, so $\dist(x,y)\le2$. Starting from $\NetT$ of \cref{cor:onestep}, add the relays $h_{w_1},\dots,h_{w_k}$ one after another as in \cref{prop:relay}, deleting the arcs $a_{x,3}\to b_{y,1}$ of $P_l\times N_l$; by the proposition the maximum flow value does not change. A pair $(x,y)\in P\times N$ at distance $2$ is not adjacent and has a common neighbour $w$, which is adjacent to $x\in P$ and $y\in N$, hence a witness, with $(x,y)\in P_l\times N_l$ for $w=w_l$; so every arc $a_{x,3}\to b_{y,1}$ of $\NetT$ is deleted, and the result is $\NetW$. \Cref{cor:onestep} gives $\W=3A-F^{\rm w}$.

(ii) Let $\Sigma=\sum_{x\in P\cup N}d_x\le\li$. Every vertex of $P\cup N\subseteq S$ has degree at least $1$, so $|P|+|N|\le\Sigma$. The relay arcs are in bijection with the pairs $(x,w)$, $x\in P\cup N$, $w\in\Wit\cap\N(x)$, i.e.\ with some adjacency entries of $P\cup N$: at most $\Sigma$ of them. Every witness has at least one relay arc in and one out, so $|\Wit|\le\Sigma/2$. An adjacent pair $(x,y)\in P\times N$ uses the entry $y$ of the list of $x$ and the entry $x$ of the list of $y$, and distinct pairs use distinct entries, so $2e_1\le\Sigma$. Hence $\NetW$ has $2(|P|+|N|)+|\Wit|+2\le\frac52\Sigma+2\le3\li+2$ nodes and $2|P|+2|N|+|P|+2e_1+(\text{relay arcs})\le3\Sigma+\Sigma+\Sigma\le5\li$ arcs.

(iii) Keep integer arrays over $V$, all initialized once to $-1$, and a counter $c$ incremented for every edge; a vertex is marked in an array by writing $c$ there, so the marks of previous edges are void without clearing. Mark $\N(i)$ and scan $\N(j)$ to split the two lists into $\Ui$, $\Uj$, $\Cc$ ($O(d_i+d_j)$, no sorting needed), and compute the masses, $P$, and $N$. Mark the vertices of $P$ in a first array and those of $N$ in a second, with their local indices in two companion arrays. Scan the lists of the vertices of $P$: for an entry $w$, record the adjacent pair if $w$ is marked in the second array, and mark $w$ in a third array (``seen from $P$''). Scan the lists of the vertices of $N$: an entry $w$ marked in the third array is a witness; if it is not yet marked in a fourth array, mark it there, give it the next index (in a companion array) and a node $h_w$; append the arc $h_w\to b_{y,1}$. Scan the lists of $P$ once more and append $a_{x,3}\to h_w$ for every entry $w$ marked in the fourth array. Every list is read at most twice, and the arcs, as many as bounded in (ii), are put in compressed adjacency form by one counting pass.
\end{proof}

\begin{lemma}[Graph min-cost flow]
\label{lem:beckmann}
For one-step measures, $\W(m_i,m_j)=\min\big\{\sum_{(x,y)}f_{xy}:f\ge0,\ \sum_y f_{xy}-\sum_y f_{yx}=\mu(x)\ \forall x\big\}$, where $f$ ranges over flows on both orientations of the edge $ij$ and of the edges with an endpoint in $P\cup N$, of which there are at most $\li+1$.
\end{lemma}

\begin{proof}
On the whole graph, the right-hand side (with all edges) is $\W$ for the hop metric: a flow decomposes into paths and cycles, and paths from $P$ to $N$ cost at least the distance, while an optimal plan routed along shortest paths gives a flow of cost $\W$ \citep{beckmann_continuous_1952,peyre_computational_2019}. Restricting to fewer edges can only increase the minimum. An optimal plan pairs vertices $x\in P$ with vertices $y\in N$, and each pair can be joined by a shortest path of one of the forms of \cref{lem:distances}: the edge $xy$; $x,w,y$ for a common neighbour $w$ (possibly $i$ or $j$); or $x,i,j,y$. Every edge of these paths has an endpoint in $P\cup N$, except $ij$ in the last form, so routing the plan along them gives a feasible flow of cost $\W$ on the restricted edges. The bound on their number holds because every such edge other than $ij$ is an adjacency entry of $P\cup N$.
\end{proof}

\subsection{\texorpdfstring{\Cref{prop:dasgupta}}{Proposition 8}}

\begin{proof}
The graph $G_H$ of \citet[Proposition~2]{dasgupta_computing_2023} has the edge $(i,j)$, $i$ adjacent to $L=\{x_1,\dots,x_n\}$, $j$ adjacent to $R=\{y_1,\dots,y_n\}$, an edge $x_ay_b$ for every pair of weight $1$, and a new vertex $z_{ab}$ adjacent to $x_a$ and $y_b$ for every pair of weight $2$. Then $d_i=d_j=n+1$, $\Cc=\emptyset$, and for the lazy walk $\mu$ vanishes on $\{i,j\}$, so $P=L$, $N=R$, every vertex with mass $1/(n+2)$, and $\dist(x_a,y_b)=w(x_a,y_b)$ by \cref{lem:distances}. The witnesses are exactly the vertices $z_{ab}$ ($i$ and $j$ are adjacent to only one side, and no two vertices of $L$, or of $R$, are adjacent). (i) Counting as in \cref{thm:witness}: $4n+2$ nodes of $\NetT$ plus $e_2$ relays; $5n$ source, sink, and chain arcs, $2e_1$ adjacency arcs, and two arcs per relay. The local input is $\li=2(n+1)+2n+2e_1+2e_2$. Since $\W=M(H)/(n+2)$ \citep[Section~2.2]{dasgupta_computing_2023}, $M(H)=3n-F^{\rm w}$ with unit masses. (ii) In the proofs of their Theorem~4, $e_2=0$ and the pairs of weight $1$ form a matching (possibly empty). Then, besides the chain $a_{x,3}\to a_{x,2}$, the arcs out of $a_{x,2}$ and $a_{x,3}$ go only to the column $y$ matched with $x$, and the network consists of the paths $\s\to a_{x,\tau}\to b_{y,\tau-1}\to\tsink$ ($\tau=2,3$) of the matched pairs, the chains, and rows and columns without arcs between them. If there is no matched pair, $\tsink$ is not reachable and Dinic's algorithm stops after its first search. Otherwise, in its first phase the chains join nodes of the same level and are not used; the blocking flow saturates every source arc of a matched row along the paths of length $3$, after which the only unsaturated source arcs lead to unmatched rows, from which no column can be reached. The phase takes time linear in the network, so with (i) the exact curvature takes $O(n)$ time.
\end{proof}

\subsection{\texorpdfstring{\Cref{thm:matching}}{Theorem 9}}

\begin{proof}
In $G_B$, $d_i=d_j=n+1$ and $\Cc=\emptyset$, so for the lazy walk $\mu$ vanishes on $\{i,j\}$, $P=L$, $N=R$, and every vertex of $L\cup R$ has mass $1/(n+2)$. A vertex of $L$ is adjacent only to $i$ and to vertices of $R$, and a vertex of $R$ only to $j$ and to vertices of $L$, so no pair of $L\times R$ has a common neighbour, and by \cref{lem:distances} $\dist(x,y)$ is $1$ if $xy\in E_B$ and $3$ otherwise. With equal masses, the transport problem is an assignment problem, which has an optimal solution that is $1/(n+2)$ times a perfect matching of $L\times R$ \citep{ahuja_network_1993}. A perfect matching with $k$ pairs in $E_B$ costs $k+3(n-k)=3n-2k$, and every matching of $B$ extends to a perfect matching of $L\times R$, so the minimum is $3n-2\alpha'(B)$. The local input is $d_i+d_j+\sum_{x\in L\cup R}d_x=2(n+1)+2n+2|E_B|$, and $G_B$ is built in $O(n+|E_B|)$ time.
\end{proof}

\subsection{\texorpdfstring{\Cref{prop:phases}}{Proposition 10}}

\begin{proof}
Multiplying all capacities by the same factor does not change the phases of Dinic's algorithm (the level graphs and the choices are the same, and every flow is multiplied), so we may use the smallest integer scaling: flows are integral, and every phase increases the flow by at least $1$. Call the nodes $h_w$ and $H_1,\dots,H_4$ \emph{relays}. Relays are joined only to $a$-nodes and $b$-nodes, and $\s$ and $\tsink$ only to $a$-nodes and $b$-nodes, so no two relays are adjacent, also in residual networks. Under any feasible flow, the flow entering a node $v$ is at most $T(v)$, where $T(a_{x,3})=\mu(x)$ (its only entering arc is from $\s$; for a merged class of rows, $\mu$ of the class), $T(a_{x,2})=2\mu(x)$ (its entering arcs are from $\s$ and from $a_{x,3}$), $T(b_{y,1})=\nu(y)$, and $T(b_{y,2})=2\nu(y)$ (their leaving arcs go to $\tsink$ and, for $b_{y,2}$ if the chain is kept, to $b_{y,1}$); these sum to at most $6A$.

Let $f$ be the flow at the start of a phase, $f^*$ a maximum flow, and $\tilde M$ the difference of their values. Then $g=f^*-f$ is a flow of value $\tilde M$ in the residual network of $f$ that enters each node $v$ with at most $2T(v)$ units: an arc into $v$ carries at most $f^*$ forward, and an arc out of $v$ at most $f$ backward. Decompose $g$ into $\tilde M$ unit $\s$--$\tsink$ paths and cycles. Each path has length at least the current distance $l$ from $\s$ to $\tsink$, so at least $(l-1)/2$ of its intermediate nodes are not relays, and each node $v$ that is not a relay lies on at most $2T(v)$ of the paths. Hence $\tilde M\,(l-1)/2\le 12A$, that is, $l\le 24A/\tilde M+1$. The distance increases from phase to phase \citep{even_network_1975}; while $\tilde M\ge\sqrt{24A}$, $l\le\sqrt{24A}+1$, so there are at most $\sqrt{24A}+1$ such phases, and afterwards at most $\sqrt{24A}$, since each increases the flow by at least $1$.

A phase takes $O(m)$ time for the breadth-first search, the current-arc pointers, and the dead ends, plus the length of each augmenting path. Each augmentation sends at least one unit, so when $r$ units remain it follows a path of length at most $24A/r+1$; the flow is at most $2A$, the capacity out of $\s$, so the augmenting paths have total length $O(A\log A)$, and the time is $O(\sqrt{A}\,m+A\log A)$. For the lazy walk with $d_i=d_j$, $\mu$ vanishes on $\{i,j\}\cup\Cc$ and every vertex of $\Ui$ and $\Uj$ has the same mass, so $A=|\Ui|$; with $m\le5\li$ (\cref{thm:witness}) and $|\Ui|\log|\Ui|\le\li\sqrt{|\Ui|}$, the time is $O(\li\sqrt{|\Ui|})$.
\end{proof}

\subsection{\texorpdfstring{\Cref{lem:locality}}{Lemma 11}}

\begin{proof}
Let $(i,j)$ be an edge of $G'$ with $i,j\notin U$. No changed pair contains $i$ or $j$ (each $a_l,b_l\in U$), so $(i,j)\in E$, $\N(i)$ and $\N(j)$ are the same in $G$ and $G'$, and so are $S$, $m_i$, and $m_j$. Moreover, no $a_l$ or $b_l$ lies in $S$: if $a_l\in\Nc{i}$ in $G'$, then $i\in\Nc{a_l}\subseteq U$, and if $a_l\in\Nc{i}$ in $G$ but not in $G'$, then $(i,a_l)$ is a changed pair, which contains $i$; the same holds for $j$ and for $b_l$. By \cref{lem:distances} and \cref{thm:layered}, $\kOR(i,j)$ is determined by $m_i$, $m_j$, and, for pairs $x,y\in S$, whether $xy\in E$ and whether $x$ and $y$ have a common neighbour. A changed pair $(a_l,b_l)$ alters the first relation only for $\{x,y\}=\{a_l,b_l\}$ and the second only for pairs containing $a_l$ or $b_l$ (a path $x,w,y$ uses the pairs $\{x,w\}$ and $\{w,y\}$), and neither $a_l$ nor $b_l$ is in $S$. Hence $\kOR(i,j)$ is the same in $G$ and $G'$.
\end{proof}

\section{Implementation}
\label{app:impl}

\begin{algorithm}[H]
\caption{Exact curvature of one edge $(i,j)$, $d_i\ge d_j$ (one-step measures)}
\label{alg:solver}
\begin{algorithmic}
\STATE Merge the sorted lists $\N(i)$ and $\N(j)$ into $\Ui$, $\Uj$, $\Cc$; rows $\Nc{i}$, columns $\Nc{j}$
\STATE Integer masses $\Lambda\,m_i$, $\Lambda\,m_j$ (per block, or per vertex for weighted walks); net masses $\mu$ on rows, $\nu$ on columns
\IF{$A=0$}
\RETURN $\kOR=1$
\ENDIF
\STATE $\theta\gets 4\big(\sum_{x\in\Nc{i}}d_x+\sum_{y\in\Nc{j}}d_y\big)$ \COMMENT{bound on the bitset work, in words}
\STATE For every vertex $w$ adjacent to a column: bitset $\mathrm{Col}(w)$ of its column neighbours; if this needs more than $\theta$ words, \textbf{go to} the last line
\FOR{every row $x$}
\STATE $\mathrm{Adj}(x)\gets \mathrm{Col}(x)$; $\mathrm{Cov}(x)\gets\bigcup_{w\in\N(x)}\mathrm{Col}(w)$, stopping once $\mathrm{Cov}(x)\supseteq\Uj$; if more than $\theta$ words have been combined and most rows so far are not full, \textbf{go to} the last line
\ENDFOR
\IF{every row of $\Ui$ is full}
\STATE Merge rows with equal $\mathrm{Adj}\cap N$; $F_1\gets$ max-flow of the one-layer network; \RETURN $\kOR=1-(2A-F_1)/\Lambda$
\ENDIF
\STATE Merge rows with equal keys $(\mathrm{Adj}\cap N,\mathrm{Cov},\text{block})$; build $\NetT$ with relays $H_1,\dots,H_4$ and, for the remaining pairs at distance $2$, witness relays if an upper bound on their arcs is below a quarter of the explicit arcs, and explicit arcs otherwise
\IF{the network has at least $300$ arcs}
\STATE $F_1\gets$ max-flow of the one-layer network, loaded on both layers \COMMENT{warm start}
\ELSE
\STATE $F_1\gets 0$
\ENDIF
\STATE $F\gets 2F_1+{}$Dinic augmentation on the residual network; \RETURN $\kOR=1-(3A-F)/\Lambda$
\STATE \textit{(bitsets abandoned)} Build $\NetW$ of \cref{thm:witness} with twin merging; warm start and Dinic as above; \RETURN $\kOR=1-(3A-F^{\rm w})/\Lambda$
\end{algorithmic}
\end{algorithm}

\paragraph{Masses.} With $\varrho=\max\{d_i,d_j\}$, the integer masses on the blocks $\{i\},\{j\},\Cc,\Ui,\Uj$ are, for $m_i$ and $m_j$ respectively: lazy, $\Lambda=(d_i+1)(d_j+1)$, $m_i\equiv d_j+1$ on $\Nc{i}$ and $m_j\equiv d_i+1$ on $\Nc{j}$; non-lazy, $\Lambda=d_id_j$, $m_i\equiv d_j$ on $\N(i)$ and $m_j\equiv d_i$ on $\N(j)$; idleness $a/b$, $\Lambda=b\,d_id_j$, $m_i(i)=a\,d_id_j$ and $m_i\equiv(b-a)d_j$ on $\N(i)$ (symmetrically for $j$); LLY, idleness $1/(\varrho+1)$ and $\kOR_{\rm LLY}=(\varrho+1)\kOR/\varrho$. For a weighted walk with integer edge weights $w$ and rational idleness $a_u/b_u$, $m_u(u)=a_u/b_u$ and $m_u(x)=(1-a_u/b_u)\,w_{ux}/\sum_{z}w_{uz}$; per-vertex denominators are reduced once, and $\Lambda$ is the least common multiple of those of $i$ and $j$.

\paragraph{Relays in detail.} Rows are $\Ui$ followed by $i$, $j$, and $\Cc$; columns likewise. $H_1$ receives an arc from $a_{x,3}$ of every row class and sends arcs to $b_{y,1}$ for the columns of $N$ in $\{i,j\}\cup\Cc$; $H_2$ receives from the row classes of $P$ in $\{i,j\}\cup\Cc$ and sends to all columns of $N$; $H_3$ receives from the full row classes and sends to the columns of $N\cap\Uj$; $H_4$ receives from the non-full row classes of $\Ui$ and sends to the full columns. A relay is created only if both of its sides are non-empty. Witness relays inside the bitset network serve only the remaining pairs: a middle vertex $w$ gets a relay from the non-full row classes of $\Ui$ adjacent to it to the non-full columns of $N\cap\Uj$ adjacent to it, and only if both sides have more than one element (otherwise its pairs get direct arcs).

\paragraph{Witness network.} Rows of $\NetW$ with the same adjacent columns and the same witnesses are merged; their keys are the sorted codes $2v+[v\text{ is a witness}]$ of their relevant neighbours, which come out of the sorted adjacency lists in order. A witness that reaches a single row class or a single column is replaced by direct arcs. The masses and the orientation are those of the bitset kernel, so the exact fractions are identical (tested for six measures and every flag).

\paragraph{Execution.} Edges are split cyclically into $8T$ chunks over $T$ threads (Numba \texttt{prange}); each thread owns growable integer arenas, so no memory is allocated per edge, and the workspace beyond the CSR graph is that of the largest local network per thread, plus the arrays of the witness network, allocated by a thread when it first hands an edge over to it. The warm start is used on networks with at least $300$ arcs.

\begin{figure}
\centering
\includegraphics[width=\textwidth]{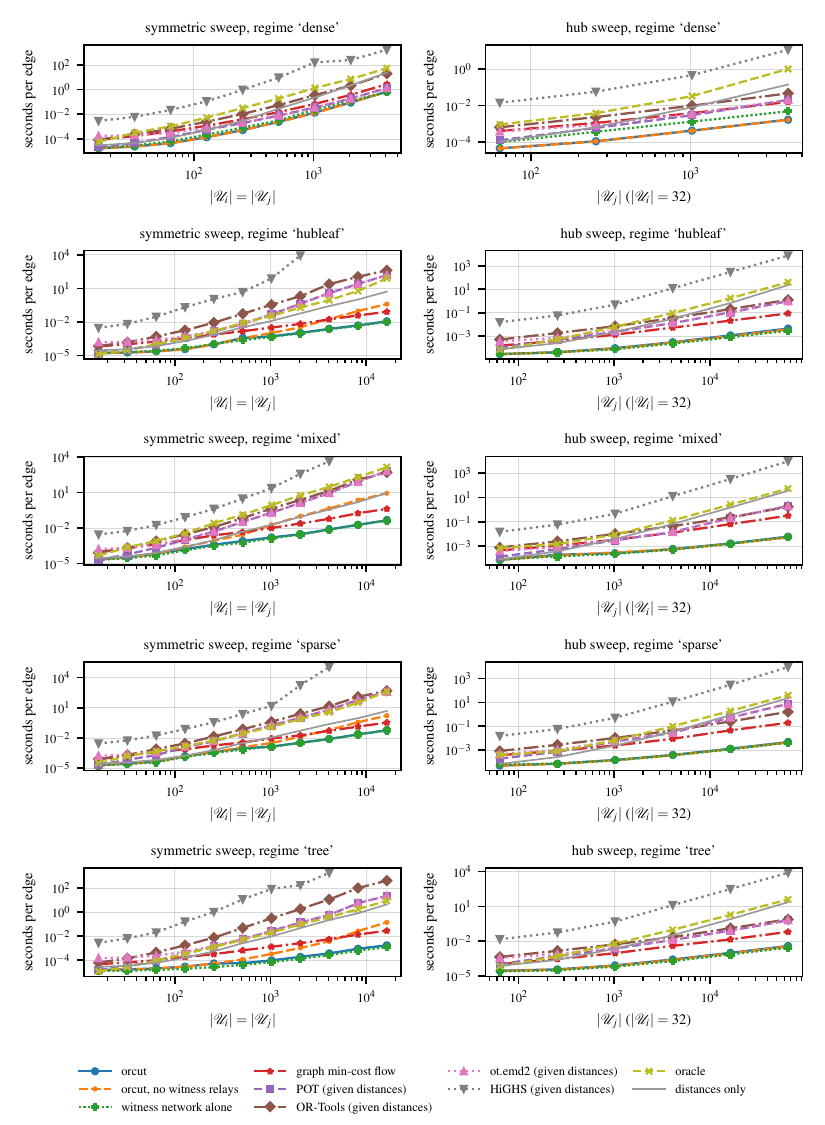}
\caption{Time per edge on controlled neighbourhoods, all five regimes and all methods (as \cref{fig:per-edge}).
The \emph{oracle} is the independent implementation on explicit distance matrices of \cref{app:impl}.}
\label{fig:per-edge-all}
\end{figure}

\paragraph{Interface and tests.} The package exposes \texttt{curvature(graph, measure)}, which returns exact fractions and per-edge diagnostics on request and lets every step of \cref{sec:solver} be switched off for ablations; \texttt{certify(graph, edge, measure)} (\cref{prop:cert}), which returns an optimal plan and an optimal potential, verifies both exactly, and can recheck them from scratch with breadth-first distances; \texttt{ball\_curvature} and \texttt{walk\_curvature} (\cref{thm:layered} with $D=2r+1$); and \texttt{affected\_edges} (\cref{lem:locality}). The tests check \texttt{affected\_edges} against full recomputation after random insertions and deletions, and compare all kernels with each other, with linear programming, and with the \emph{oracle} of \cref{tab:ablation} and \cref{fig:per-edge-all}, an independent implementation that builds the full local distance matrix, merges twins, and solves the three-level closure with its own distance and maximum-flow code.

\section{Experimental Setup}
\label{app:setup}

\paragraph{Hardware and software.} Nodes of a Slurm cluster with two Intel Xeon Gold 6148 (20 cores each, 2.4\,GHz), two Xeon Gold 6248 (20 cores, 2.5\,GHz), two Xeon Gold 6230 (20 cores, 2.1\,GHz), two AMD EPYC 7713 (64 cores, 2.0\,GHz), or four Xeon Gold 6148, and \res{EThreeThreads} threads in every job. A step stopped by its time limit resumed on whichever of these nodes was free, so the rows of a table, and the chunks of its largest rows, can come from different processors (those of Orkut from two). The solver and every experiment are run with Python \res{EnvPython}, NumPy \res{EnvNumpy}, Numba \res{EnvNumba}, \texttt{POT} \res{EnvPot}, \texttt{OR-Tools} \res{EnvOrtools}, and NetworkX \res{EnvNetworkx} \citep{hagberg_networkx_2008}; GraphRicciCurvature 0.6.1 ran with SciPy 1.13 and NetworKit \citep{staudt_networkit_2016} in a separate environment.

\paragraph{Calibration across node types.}
To measure how processor differences affect times and ratios, we timed \res{ECalibGraphs} again on one node of each of \res{ECalibNodeTypes} types (\cref{tab:calib}). Across node types, our times vary by up to \res{ECalibTimeSpreadMax}, \texttt{POT}'s estimates by up to \res{ECalibPotSpreadMax}, and their ratio by up to \res{ECalibRatioSpreadMax}. The spread comes mainly from the AMD EPYC node, where our solver runs faster and yields higher speedup ratios; the two Intel Xeon processors differ little from each other, and each graph's solve-only ratio varies little across node types (\cref{tab:calib}). E3's comparisons with \texttt{POT} are unaffected: every row except Orkut was measured in one run on a node with two Xeon Gold 6248, and Orkut's chunks ran on that node and on the one with four Xeon Gold 6148 that also gave its \texttt{POT} estimate, the two Intel processors of \cref{tab:calib}, which yield the more conservative speedups.

\begin{table}
\caption{The same graphs timed on one node of each type, median of three runs, \res{EThreeThreads} threads, estimate of \texttt{POT} samples $20$ edges per degree stratum. Ratios: \texttt{POT}'s estimate over our time, with its distance matrices and for its solve alone.}
\label{tab:calib}
\centering
\small
\begin{tabular}{@{}llrrcc@{}}
\toprule
Processor & Graph & ours & \texttt{POT} (est.) & ratio & solve only \\
\midrule
\resrows{calib}
\bottomrule
\end{tabular}
\end{table}

\paragraph{The node behind each row.}
The logs record when each row was written, and the scheduler records which job ran on which node, so every row can be traced to its processor (\cref{tab:provenance}). Of the \res{EProvRows} rows traced, \res{EProvOpen} (all from E3's random graphs) were written while workers ran on two node types at once. A row's own comparisons come from one run, except for E3's chunked rows above. Comparisons across rows of the same graph mix processors only in E5, on \res{EProvFiveWholeMixed} of \res{EProvFiveWholeGraphs} graphs against GRC and \res{EProvFiveSinkhornMixed} of \res{EProvFiveSinkhornGraphs} against Sinkhorn; in E4 and E6, each graph's rows come from one processor.

\begin{table}
\caption{The processors behind each experiment, traced from the logs and the scheduler's records: the number of rows each processor wrote. Not determined (ND): written while workers of the step ran on two types of node.}
\label{tab:provenance}
\centering
\small
\begin{tabular}{@{}lp{0.66\textwidth}@{}}
\toprule
Experiment & Processors (rows) \\
\midrule
\resrows{provenance}
\bottomrule
\end{tabular}
\end{table}

\paragraph{Graphs.} Every real network is restricted to its largest connected component and made simple and undirected: the 20 networks of \cref{tab:datasets} and five small ones that ship with the code, Karate \citep{zachary_karate_1977}, Jazz \citep{gleiser_community_2003}, Yeast \citep{milo_motifs_2002}, Power Grid \citep{watts_collective_1998}, and arXiv HepPh \citep{gehrke_arxiv_2003}. The random families, all seeded, are \citet[][ER]{erdos_random_1959} $G(n,m)$, \citet[][BA]{barabasi_emergence_1999}, random geometric graphs in the unit square (RGG, \citealp{penrose_random_2003}), \citet[][WS]{watts_collective_1998} with rewiring probability $0.2$, and \citet{chung_random_2002} graphs with power-law exponents $2.1$ and $2.5$ (CL2.1, CL2.5); a random graph is written family($n$, average degree), as in ER(2k, 10) for $n=2{,}000$ and average degree $10$.

\begin{figure}
\centering
\includegraphics[width=\textwidth]{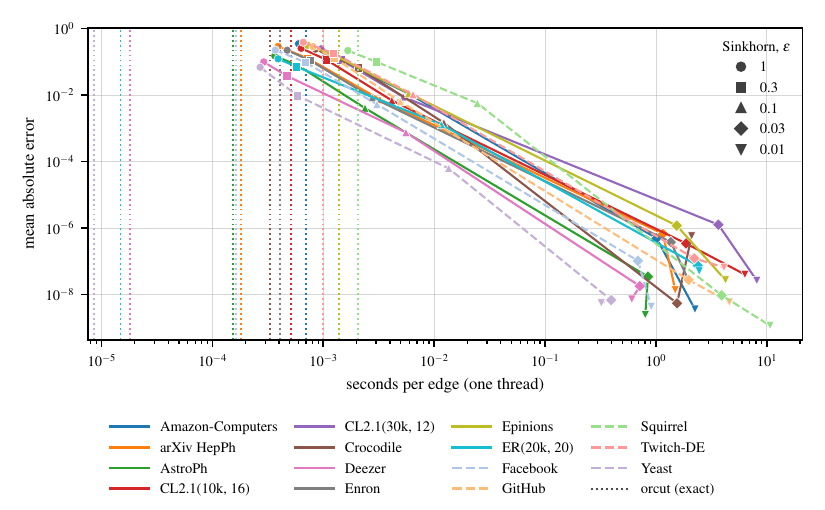}
\caption{Mean absolute error of the curvature at idleness $1/2$ against the mean time per edge, one thread, for each graph (colour and line style) and each regularization $\varepsilon$ (marker); Sinkhorn is given the distances. The dotted vertical lines mark the mean time per edge of the exact solver on each graph, in the colour of that graph; its error is zero, so it has no point on the logarithmic axis.}
\label{fig:tradeoff}
\end{figure}

E1 uses the five small networks, ER(2k, 10), BA(2k, 10), WS(2k, 10), RGG(2k, 12), CL2.1(3k, 8), CL2.5(3k, 8), and eight canonical graphs: $K_{20}$, $Q_6$, the $20\times20$ grid, the binary tree of depth~8, $C_{100}$, $K_{1,200}$, the wheel with 50 rim vertices, and $K_{10,30}$. E2 uses the small networks (except Karate) and the 14 networks of \cref{tab:datasets} with fewer than $10^6$ edges. E3 uses all 20 networks of \cref{tab:datasets} and every random family, at average degree $16$ for $n=10^3,3\cdot10^3,\dots,10^7$ (CL2.1 up to $3\cdot10^6$) and at average degree $2.1\,n^{0.4}$ for $n$ up to $10^6$ (ER and BA up to $3\cdot10^5$, CL2.1 and CL2.5 up to $10^5$), one seed each. E4 uses ER(20k, 20), BA(20k, 10), RGG(20k, 20), CL2.1(50k, 10), arXiv HepPh, GitHub, Twitch-DE, Squirrel, Epinions, and DBLP; E5 uses the \res{EFiveGraphs} graphs of \cref{fig:tradeoff}; and E6 uses arXiv HepPh, Deezer, Twitch-DE, GitHub, Amazon-Computers, Epinions, DBLP, YouTube, RGG(100k, 20), and CL2.1(30k, 12) for the other one-step measures, and Power Grid, Yeast, Jazz, ER(5k, 8), BA(5k, 6), WS(5k, 8), RGG(5k, 10), and CL2.5(5k, 6) for balls and multi-step walks.

\paragraph{Controlled edges.} A controlled edge $(i,j)$ of E2 has $|\Ui|=a$, $|\Uj|=b$, and $t$ common neighbours. The symmetric sweep has $a=b=s$ and $t=\max\{1,\lfloor s/8\rfloor\}$ for $s=2^4,2^5,\dots,2^{14}$, and the hub sweep has $a=32$, $t=4$, and $b=s\in\{2^6,2^8,\dots,2^{16}\}$; each size has three seeds. Pairs at distance $1$ are \emph{cross edges}, edges between $\Ui$ and $\Uj$, and pairs at distance $2$ come from \emph{connectors}, extra vertices adjacent to random vertices of $\Ui$ and $\Uj$. Eight connectors, each adjacent to a fraction $f$ of $\Ui$ and of $\Uj$, give about $1-(1-f^2)^8$ of the pairs a common neighbour, and $f$ is chosen to make this share $q$. 
Cross edges are $\mathrm{Bin}(ab,p)$ pairs of $\Ui\times\Uj$ drawn uniformly with replacement, so that a pair is a cross edge with probability about $1-e^{-p}$. The five regimes are \emph{tree}, with no cross edges and no connectors; \emph{sparse}, with $p=2/b$ and $q=0.05$; \emph{mixed}, with $p=2/b$ and $q=1/2$ (\cref{fig:per-edge}); \emph{dense}, with $p=0.2$ (about $18\%$ of the pairs) and one connector adjacent to all of $\Ui\cup\Uj$, so that no pair is at distance $3$ (only for $s\le2^{12}$, beyond which the cross edges are too many to generate); and \emph{hubleaf}, in which all but $2\%$ of $\Uj$ are leaves of $j$, and hence twins, while the remaining $2\%$ carry the structure of \emph{mixed}.

\paragraph{Lower-bound instances.} The families of \cref{tab:dasgupta} have an edge $(i,j)$ with $i$ adjacent to $L$, $j$ adjacent to $R$, and $|L|=|R|=n$: \emph{single}, one random adjacent pair of $L\times R$ (the instances of Theorem~4(a) of \citealp{dasgupta_computing_2023}); \emph{matching}, a random perfect matching (their Theorem~4(b)); \emph{bounded}, two random partners at distance $1$ and two at distance $2$ per vertex of $L$, each pair at distance $2$ with its own middle vertex (graphs $G_H$ with bounded weighted degrees); and \emph{blocks}, $\mathrm{round}(\sqrt n)$ shared middle vertices, each adjacent to $\mathrm{round}(\sqrt n)$ random vertices of $L$ and of $R$; $n=2^8,2^{10},\dots,2^{20}$, with two seeds.

\paragraph{Timing.} Per-edge times are the fastest of several repetitions of a batch, after compilation; whole-graph times are medians of three runs (one above $10^7$ edges). Single-threaded comparisons fix one thread for every method. \texttt{POT}'s network simplex is called through its C++ entry point with an iteration limit high enough never to bind; \texttt{ot.emd2} is called as GRC calls it. Whole-graph \texttt{POT} times are estimated as described in \cref{sec:experiments}. Graph min-cost flow (\cref{lem:beckmann}) uses \texttt{OR-Tools}' \texttt{SimpleMinCostFlow} with integer masses on both orientations of the edge $ij$ and of the edges with an endpoint in $P\cup N$, built by Numba \citep{lam_numba_2015} code that is included in its time.

\paragraph{GRC.} Version 0.6.1, called as a user would, via \texttt{OllivierRicci} and its method \texttt{compute\_\allowbreak ricci\_\allowbreak curvature\_\allowbreak edges}, with \res{EThreeThreads} processes, all-pairs shortest paths up to $2\times10^4$ vertices and pairwise shortest paths beyond (GRC builds the matrix from $n^2$ floats, $\approx48n^2$ bytes at its peak), and no time limit; \texttt{ATD} always builds this matrix and is not run where it does not fit in the job's memory. Methods: the default (\texttt{OTDSinkhornMix}, \texttt{nbr\_topk}${}=3000$), \texttt{OTD}, \texttt{OTD} without truncation, \texttt{Sinkhorn} (regularization $0.1$), and \texttt{ATD}, all at $\alpha=1/2$.\footnote{\texttt{OTD} is \texttt{ot.emd2}, exact up to \texttt{POT}'s default cap of $100{,}000$ network-simplex iterations: the \res{EFiveGrcOtdFullMaxErr} in \cref{tab:approx} is due to the size of GitHub ($6.7\times10^7$ entries). ATD is GRC's average transportation distance, and OTDSinkhornMix uses OTD unless both endpoints have degree above $2{,}000$, and Sinkhorn otherwise.}

\begin{table}
\caption{Real networks of E3 (largest connected component). All except Amazon-Photo, Amazon-Computers, and Coauthor-Physics come from \cite{leskovec_snapnets_2014}; the last column gives each network's original source.}
\label{tab:datasets}
\centering
\small
\setlength{\tabcolsep}{4pt}
\begin{tabular}{@{}lrrrrl@{}}
\toprule
Graph & $n$ & $m$ & $d_{\max}$ & mean degree & source \\
\midrule
\resrows{datasets}
\bottomrule
\end{tabular}
\end{table}

\section{Additional Results}
\label{app:results}
\label{app:more}

\begin{table}
\caption{\texttt{OR-Tools}: edges compared as fractions / mismatches. Kernels: mismatches with the oracle and the layered closure (lazy, non-lazy, and LLY curvatures). \texttt{POT}: largest absolute difference. Certificates: checked / failed.}
\label{tab:exactness}
\centering
\small
\begin{tabular}{@{}lrrrrrr@{}}
\toprule
Measure & graphs & edges & \texttt{OR-Tools} & kernels & \texttt{POT} & certificates \\
\midrule
\resrows{exactness}
\bottomrule
\end{tabular}
\end{table}

\begin{table}
\caption{Network size and runtime at the largest symmetric size. Input: $\li^S$; arcs: counts for explicit distance-$2$ pairs, the solver, and $\NetW$; exponent: fitted scaling exponent of arcs vs.\ $\li^S$; time / solver: runtime relative to the solver without witness relays (also disabling the bitset cap and $\NetW$ fallback), for $\NetW$ alone, and for graph min-cost flow.}
\label{tab:witness}
\centering
\small
\setlength{\tabcolsep}{3.5pt}
\begin{tabular}{@{}lrrrrrrrrrr@{}}
\toprule
& & & \multicolumn{3}{c}{arcs} & \multicolumn{2}{c}{exponent} & \multicolumn{3}{c}{time / solver} \\
\cmidrule(lr){4-6}\cmidrule(lr){7-8}\cmidrule(l){9-11}
Regime & $s$ & $\li^S$ & explicit & solver & witness & explicit & witness & \begin{tabular}[b]{@{}r@{}}no\\witn.\end{tabular} & witness & \begin{tabular}[b]{@{}r@{}}graph\\flow\end{tabular} \\
\midrule
\resrows{witness}
\bottomrule
\end{tabular}
\end{table}

\begin{figure}
\centering
\includegraphics[width=\textwidth]{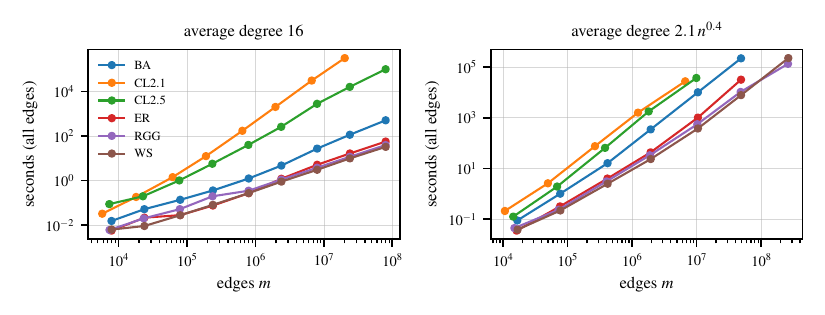}
\caption{Time for all edges of random graphs against their number of edges, with average degree $16$ (left) and $2.1\,n^{0.4}$ (right), on \res{EThreeThreads} threads; the sizes are in \cref{app:setup}.}
\label{fig:synth}
\end{figure}

\begin{table}
\centering
\begin{minipage}[t]{0.45\textwidth}
\centering
\captionof{table}{Exponent $k$ of time $\propto m^k$ on random families, fitted by least squares on logarithms over the sizes of \cref{app:setup}.}
\label{tab:synth}
\small
\setlength{\tabcolsep}{4pt}
\begin{tabular}{@{}llrr@{}}
\toprule
Family & degree & largest $m$ & $k$ \\
\midrule
\resrows{synth}
\bottomrule
\end{tabular}
\end{minipage}\hfill
\begin{minipage}[t]{0.52\textwidth}
\centering
\captionof{table}{Whole-graph time relative to the lazy walk (median and maximum over graphs).}
\label{tab:measures}
\small
\begin{tabular}{@{}lrr@{}}
\toprule
Measure & median & max \\
\midrule
\resrows{measures}
\bottomrule
\end{tabular}

\vspace{1em}

\captionof{table}{Larger supports, layered network with $D=2r+1$ (all values certified on every graph). \texttt{POT}/ours: estimated single-thread time ratio (median, range over graphs); max diff.: largest difference to \texttt{POT}.}
\label{tab:radius}
\small
\setlength{\tabcolsep}{3.5pt}
\begin{tabular}{@{}lrrr@{}>{\setbox0=\hbox\bgroup}l<{\egroup}@{}}
\toprule
Measure & \texttt{POT}/ours & range & max diff. \\
\midrule
\resrows{radius}
\bottomrule
\end{tabular}
\end{minipage}
\end{table}

\paragraph{Other measures (E6).}
Non-lazy, LLY, idleness $1/4$, $1/2$, and $3/4$, and weighted lazy walks cost within \res{ESixRelMax} of the lazy walk (median \res{ESixRelMedian}; \cref{tab:measures}). For uniform measures on balls of radius $2$ and $3$ and two- and three-step walks, our reference implementation of $\Net_{2r+1}$, which still builds the distances between $P$ and $N$, is a median \res{ESixBallTwoPotMedian} faster than \texttt{POT} on radius-$2$ balls and certifies every value it returns (\cref{tab:radius}); a distance-free construction for $r\ge2$ is future work.

\clearpage
\section{Use of Generative AI}
\label{app:genai}
A generative AI assistant (Claude, Anthropic) was used to help implement the experiment pipeline and to copy-edit the text. The author designed the method and the experiments, verified every statement and proof, and takes full responsibility for the content; all numbers are produced by the released code from the raw results.

\end{document}